\documentclass[conference]{IEEEtran}
\IEEEoverridecommandlockouts
\usepackage{cite}
\usepackage{amsmath,amssymb,amsfonts}
\usepackage{graphicx}
\usepackage{textcomp}
\usepackage{booktabs}
\usepackage{multirow}
\usepackage[most]{tcolorbox}
\usepackage{algorithm}
\usepackage{algpseudocode}
\usepackage{subcaption}
\usepackage{algorithmicx}
\usepackage[table]{xcolor}
\usepackage[autostyle]{csquotes}
\usepackage{bm}
\usepackage{tabularray}
\usepackage{array}
\usepackage{adjustbox} 
\usepackage{cuted}
\usepackage{hyperref}
\usepackage{cleveref}

\usepackage{marginnote}

\usepackage{marvosym}

\usepackage[disable]{todonotes}

\newtheorem{challenge}{\bf Challenge}
\newtheorem{definition}{\bf Definition}

\newcommand{\longchainDS}{{\textsc{LongChain-DS}}}
\newcommand{\longchainQwen}{{\textsc{LongChain-Qwen}}}
\newcommand{\model}{\textsc{RedParrot}}
\newcommand{\modelRaw}{\textsc{\model{}-RAW}}
\newcommand{\modelMan}{\textsc{\model{}-MAN}}

\newcommand{\commerceCovered}{\texttt{RED-commerce-0916}}
\newcommand{\communityCovered}{\texttt{RED-community-0916}}
\newcommand{\tradingCovered}{\texttt{RED-trading-0916}}
\newcommand{\commerce}{\texttt{RED-commerce}}
\newcommand{\community}{\texttt{RED-community}}
\newcommand{\trading}{\texttt{RED-trading}}

\newcommand{\newbird}{\texttt{BIRD-DSL}}
\newcommand{\newspider}{\texttt{Spider-DSL}}

\definecolor{lg}{rgb}{0.9, 1, 0.9}
\definecolor{lr}{rgb}{1, 0.95, 0.95}
\definecolor{dr}{rgb}{0.996, 0.749, 0.753}
\definecolor{dg}{rgb}{0.710, 1.0, 0.698}

\newcounter{observationcounter}

\newcommand{\Example}[2]{

    \refstepcounter{observationcounter}\label{#1}

    \vspace{-\topsep}

    \begin{tcolorbox}[colback=gray!10, colframe=gray!50, coltitle=black, width=\columnwidth, boxrule=0.5pt, arc=1mm, boxsep=0mm, left=1.5mm, right=1.5mm, top=1.5mm, bottom=1.5mm, breakable]

    \textbf{Example \theobservationcounter}. {#2}

    \end{tcolorbox}

    \vspace{-\topsep}

}

\newtcolorbox{promptbox}[1]{
  float*,                 
  floatplacement=t,
  enhanced,
  arc=8pt,                 % 稍微调小一点圆角，更符合学术论文审美
  outer arc=8pt,
  boxrule=0.6pt,           % 细边框更显精致
  colframe=black!80,       % 标题栏颜色
  colback=white,           % 主体背景色
  coltitle=white,          % 标题文字颜色
  fonttitle=\bfseries\sffamily, 
  title={#1},              
  left=12pt,               
  right=12pt,              
  top=8pt,                
  bottom=8pt,             
  boxsep=   5pt,
  width=\textwidth,       % 确保宽度是全页宽
  title={#1}
}

\definecolor{colorInstruction}{HTML}{FFF9E5} % 浅黄
\definecolor{colorQuery}{HTML}{FFF0F0}       % 浅红
\definecolor{colorKnowledge}{HTML}{F0F7FF}   % 浅蓝
\definecolor{colorExample}{HTML}{F2FFF2}     % 浅绿

\tcbset{
    promptline/.style={
        sharp corners,
        boxrule=0pt,
        left=8pt,
        right=8pt,
        top=5pt,
        bottom=5pt,
        boxsep=0pt,
        frame hidden,
        fontupper=\sffamily\small % 使用无衬线字体，更接近图片效果
    }
}

\newcommand{\remarkbox}[1]{
\vspace{-\topsep}
\begin{tcolorbox}[colback=gray!10, colframe=gray!50, coltitle=black, width=\columnwidth, boxrule=0.5pt, arc=1mm, boxsep=0mm, left=1.5mm, right=1.5mm, top=1.5mm, bottom=1.5mm, breakable]
\textbf{Remark}.{#1}
\end{tcolorbox}
\vspace{-\topsep}
}

\def\BibTeX{{\rm B\kern-.05em{\sc i\kern-.025em b}\kern-.08em
    T\kern-.1667em\lower.7ex\hbox{E}\kern-.125emX}}
\begin{document}
% \title{\model{}: Learning to Cache Query Semantics for Accelerating NL-to-DSL Business Analytics
\title{\model{}: Accelerating NL-to-DSL for Business Analytics via Query Semantic Caching
% \title{RedTC: Utilizing Template Caching to Accelerate NL-to-DSL for Business Intelligence
\thanks{$^*$ These authors contributed equally to this work.}
\thanks{Lidan Shou is the corresponding author.}
}
% \thanks{This work was supported by the Pioneer R&D Program of Zhejiang (No. 2024C01021), "Leading Talent of Technological Innovation Program" (No. 2023R5214) of Zhejiang Province, and the collaborative project between Xiaohongshu and ZJU.}
% }

\author{
% 在您的文档导言区（\documentclass 后面），请确保包含了 amsmath 包
% \usepackage{amsmath}

    \IEEEauthorblockN{
        Tong Wang\textsuperscript{*}\textsuperscript{$\dagger$},
        Yongqin Xu\textsuperscript{*}\textsuperscript{$\dagger$},
        Jianfeng Zhang\textsuperscript{$\dagger$},
        Lingxi Cui\textsuperscript{$\dagger$}, \\
        Wenqing Wei\textsuperscript{$\ddagger$},
        Suzhou Chen\textsuperscript{$\ddagger$},
        Huan Li\textsuperscript{$\dagger$},
        Ke Chen\textsuperscript{$\dagger$},
        Lidan Shou\textsuperscript{$\dagger$}\textsuperscript{\Letter}
    }
    \IEEEauthorblockA{
        \textsuperscript{$\dagger$}State Key Laboratory of Blockchain and Data Security, Zhejiang University\quad
        \textsuperscript{$\ddagger$}Xiaohongshu \\
        % \footnote{\textsuperscript{*}These authors contributed equally to this work. \quad} 
        % \textsuperscript{$\bowtie$}Corresponding author. \\
        \{tongwang, xuyongqin, jianfeng.zhang, cuilingxi.cs, lihuan.cs, chenk, should\}@zju.edu.cn, \\
        \{zengzicheng, chensuzhou\}@xiaohongshu.com
    }
}

% \and
% \and
% \and
% \and
% \and

\maketitle

% \noindent
% \textsuperscript{*}These authors contributed equally to this work.
% \textsuperscript{$\bowtie$}Corresponding author.

\begin{abstract}

% Recently, at Xiaohongshu, business teams have exhibited a rapidly growing demand for business analytics. While Large Language Models (LLMs) have significantly advanced the capabilities of natural-language driven business analytics, prevailing approaches relying on long-chain workflows with multiple LLM calls (e.g., for question parsing, data retrieval, and data analysis) suffer from high latency and compounding error propagation. In this paper, we propose \model{}, a novel NL-to-DSL framework for business analytics based on the concept of query semantics caching. Experiments on our real enterprise datasets and adapted open-source datasets show that \model{} substantially improves efficiency without compromising the accuracy of full LLM pipelines.

Recently, at Xiaohongshu, the rapid expansion of e-commerce and advertising demands real-time business analytics with high accuracy and low latency. To meet this demand, systems typically rely on converting natural language (NL) queries into Domain-Specific Languages (DSLs) to ensure semantic consistency, validation, and portability. However, existing multi-stage LLM pipelines for this NL-to-DSL task suffer from prohibitive latency, high cost, and error propagation, rendering them unsuitable for enterprise-scale deployment. 
In this paper, we propose \model{}, a novel NL-to-DSL framework that accelerates inference via a semantic cache. Observing the high repetition and stable structural patterns in user queries, \model{} bypasses the costly pipeline by matching new requests against cached \enquote{query skeletons} (normalized structural patterns) and adapting their corresponding DSLs. Our core technical contributions include (1) an offline skeleton construction strategy, (2) an online, entity-agnostic embedding model trained via contrastive learning for robust matching, and (3) a heterogeneous Retrieval-Augmented Generation (RAG) method that integrates diverse knowledge sources to handle unseen entities.
Experiments on six real enterprise datasets from Xiaohongshu show \model{} achieves an average 3.6x speedup and an 8.26\% accuracy improvement. Furthermore, on new public benchmarks adapted from Spider and BIRD, it boosts accuracy by 34.8\%, substantially outperforming standard in-context learning baselines.

\end{abstract}

% \footnotetext{\textsuperscript{*}These authors contributed equally to this work.}


\section{Introduction}

% Compact Para 1
At leading social networking platforms like Xiaohongshu, the rapid expansion of e-commerce and advertising operations has created a heightened demand for real-time business analytics. This necessitates supporting analytical workloads characterized by frequent requirement changes, high throughput, and low latency, making natural language-driven analytics an essential solution for business end-users.

\begin{figure}[t]
\centerline{\includegraphics[width=1\columnwidth]{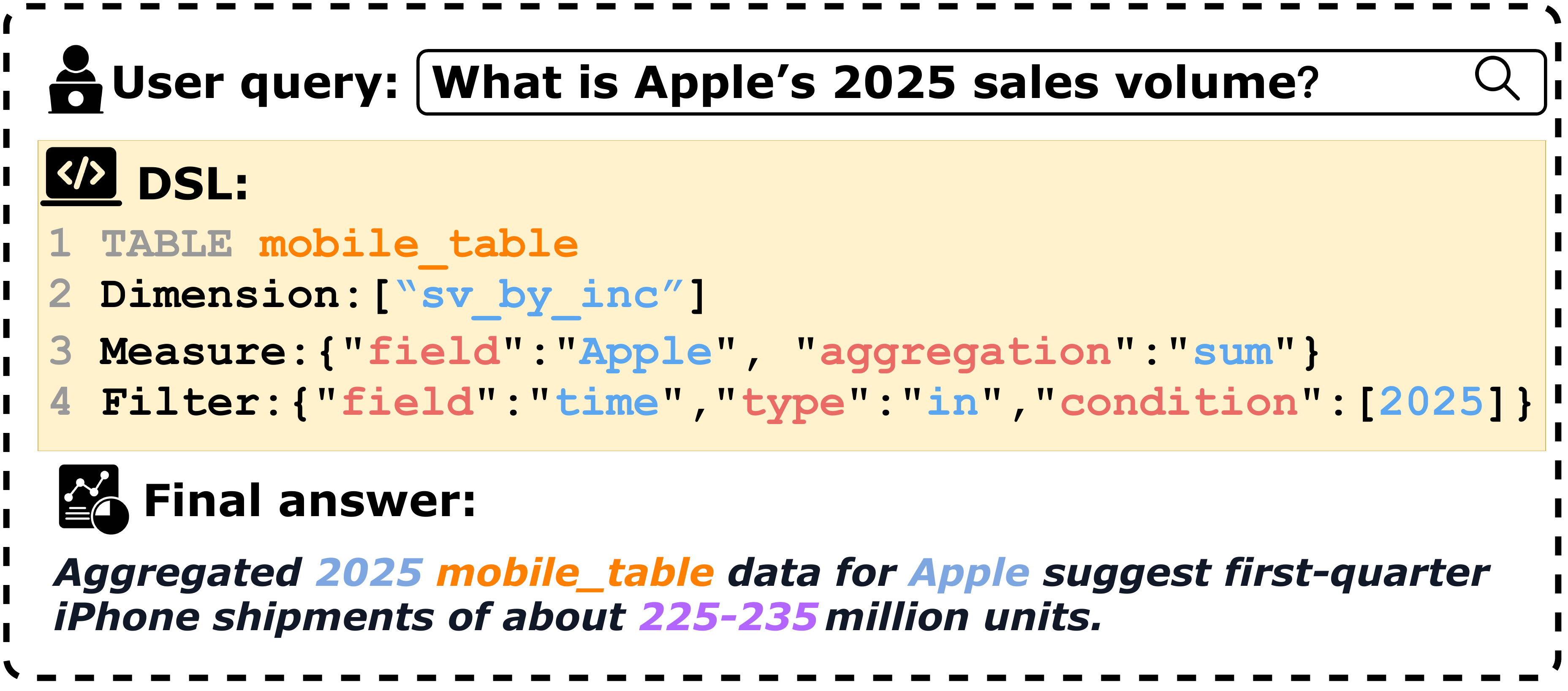}}
\caption{The NL-to-DSL translation process in business analytics context.}
\label{fig1:dsl}
\end{figure}

Recent advances in large language models (LLMs) enable Xiaohongshu to explore \emph{natural-language (NL) driven business analytics}, a paradigm that processes NL queries against a data lake to return answers as text or visualizations~\cite{zhu2024chat2query, weng2025datalab,cui2025tablecopilot}.
A common industry practice to facilitate this involves employing pre-computed wide tables to offload costly joins and unions, which operate in concert with a semantic layer defined by a domain-specific language (DSL). 
Such DSL (as exemplified in Fig.~\ref{fig1:dsl}) enforces consistent semantics, supports validation and safe evolution, ensures portability by compiling to multiple backends (e.g., SQL plans or visualizations), and allows for both direct editing and GUI-based authoring. 
Consequently, our core task is NL-to-DSL: converting personalized NL requirements into structured DSLs.

\begin{figure*}[htbp]
\centerline{\includegraphics[width=1.8\columnwidth]{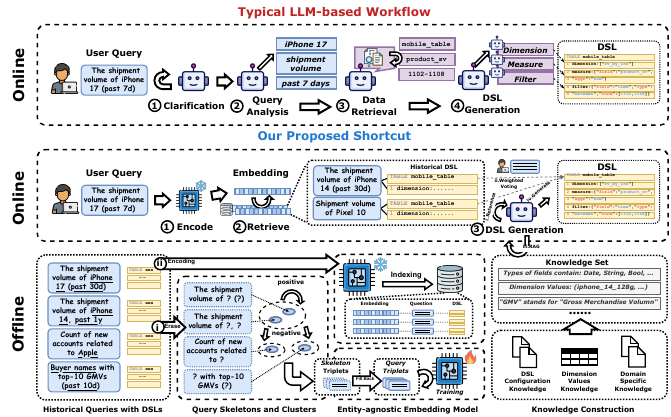}}
\caption{
% An overview of the NL2DSL procedure in a BI scenario. 
(Top) The typical agentic workflow is a long-chain solution requiring multiple LLM calls, the cost of which is analyzed in Section~\ref{section:data-analysis}. (Bottom) In contrast, our proposed \model{} leverages historical skeleton caching to create an accelerating shortcut, significantly simplifying the workflow.}
\label{fig2:main}
\vspace{-3mm}
\end{figure*}

% Para 3
Given the scarcity of labeled real-user data, which makes training-heavy end-to-end methods impractical and brittle, Xiaohongshu adopted a long-chain pipeline, as shown in Fig.~\ref{fig2:main}, which leverages the strong few-shot capability of LLMs to tackle the problem of NL-to-DSL.
This pipeline decomposes the task into four stages, namely query parsing~\cite{yuan2025cogsql,gao2024text,peng2024large}, data retrieval~\cite{balaka2025pneuma,xu2024kcmf,wang2023solo}, data analysis~\cite{hu2025lakevisage,weng2025datalab}, and validation. Each stage may require one or more LLM calls to progressively resolve semantic ambiguity, enable tighter operational control, and improve precision.
Nevertheless, the long-chain pipeline solution incurs large latency ($>$30 seconds) and cost ($>$26000 tokens per query), apart from errors that propagate across stages (see Section~\ref{2-Cost-Analysis}).

\Example{ex1}{

\textbf{Q1.} \enquote{Apple 25 sales}~\textbf{Q2.} \enquote{Huawei's sales from 23 to 25}

For a query about ``Apple 25 sales'', the target DSL can be quickly produced by adapting a cached DSL for a structurally identical query, such as ``Huawei's sales from 23 to 25'', substituting the company name and time span.
}

% Para 4
Following the deployment of our long-chain NL-to-DSL pipeline, we have amassed a corpus of real user interactions. Drawing on our experience in data management, we observe that \emph{caching and serving semantically similar queries} can substantially reduce both wall-clock latency and token usage, thereby bypassing the pipeline for most requests. In practice, user queries exhibit \textbf{high levels of repetition and stable structural patterns} (as shown in Example~\ref{ex1}), making high cache-hit rates possible.

% Para 5
Motivated by this, we advocate to cache frequently occurring queries in proper form and their DSLs as templates. 
Specifically, to enable robust matching, we distill each query into its core structural pattern, which we term a \emph{query skeleton}, by removing entity-specific and non-structural words in it. 
This normalization is critical because queries that query the same database table, despite lexical variations, often reduce to identical skeletons and thus share highly similar DSL structures. 
By retrieving semantically similar query skeletons at inference time and using their DSLs as exemplars, one can synthesize the target DSL rather than generating it from scratch. 
This not only reduces LLM usage and latency, but also improves reliability.
% 
% To meet our business needs (time and token limits), a shortcut NL-to-DSL approach using templates as a cache is the most appropriate choice. 
% 
% Therefore, we propose a shortcut NL-to-DSL framework with the following design goals:
% \begin{itemize}
    % \item \textbf{Response time:} 
    % Simplify the flow with cached templates and fewer steps to speed responses and enhance user experience.
    % \item \textbf{Token cost:}
    % Reduce the number of LLM invocations and, while preserving accuracy, minimize prompt token usage.
    % \item \textbf{Accuracy:}
    % Strive to achieve accuracy comparable to long-chain pipelines, with potential improvements.
    % \item \textbf{query Coverage:} 
    % The skeletons of template cache should span the most common query patterns and better maximize coverage.
% \end{itemize}

% Para 6 (Complete Version)
% As shown in Figure~\ref{fig:main}, we propose an end-to-end shortcut NL-to-DSL framework based on template caching, named \model{}. 
% % 
% During the offline phase, we build a template cache consisting of historical skeleton-DSL pairs that capture query patterns, and we train an entity‑agnostic embedding model via contrastive learning to filter out words unrelated to the structural pattern and derive an embedding for each query skeleton.
% % 
% At inference time, we first embed the user's query with our trained model, which allows us to retrieve the most similar skeleton-DSL pairs from our updated template cache. These selected pairs then serve as in-context demonstrations to guide the LLM's DSL generation, effectively preventing multiple LLM calls and redundant retrieval steps.

\Example{ex2}{

\textbf{Q1.} \enquote{Apple 25 sales}; \textbf{Q2.} \enquote{Huawei's sales from 23 to 25}; \textbf{Q3.} \enquote{Apple’s 2025 product catalog}.

Given the above queries, Q1 is semantically similar to both Q2 and Q3. While Q1 and Q2 differ in their mentioned entities and time span, they belong to the same structural pattern and thus map to the same table (table of annual sales). In contrast, Q1 and Q3 represent different structural patterns, eventually requiring access to distinct tables.}

Unfortunately, matching query skeletons by semantic similarity is non-trivial, as illustrated in Example~\ref{ex2}. In practice, the problem becomes even more complicated with the presence of unseen entities or attributes, jargon, and abbreviations. 
These give rise to the challenges detailed below when designing a skeleton cache solution.

% Para 7
% However, beyond the basic design goals, real-world scenarios are clearly more complex: a query may contain multiple keywords; similar queries may correspond to different tables; and even queries on the same table may be considered different due to irrelevant variables such as time constraints. This gives rise to the following main challenges:

% A new query like \enquote{Apple’s 2025 sales} can be semantically similar to historical queries such as \enquote{Huawei’s 2024 sales} and \enquote{Apple’s 2025 product catalog}. While \enquote{Apple’s 2025 sales} and \enquote{Huawei’s 2024 sales} differ in their specific entities, they conform to the same structural pattern and thus map to a single table (e.g., a table of annual sales). In contrast, \enquote{Apple’s 2025 sales} and \enquote{Apple’s 2025 product catalog} represent different query patterns, requiring access to distinct tables.
% }

\begin{challenge}\label{skeleton-challenge} 
\emph{Skeleton generation} is crucial for a performant cache.
To obtain a query skeleton, we must leverage contextual semantic information to distinguish distracting words (e.g., time expressions such as “2025” vs. “2024” in Example~\ref{ex1}) from the key terms (e.g., ``sales'').
    % How to generate can we automatically generate representative and accurate query skeletons? This is crucial both for the offline construction of a high-quality template cache and for assembling data to train the encoder via contrastive learning. Direct entity extraction is not feasible; it is necessary to leverage contextual semantic information to distinguish distractor words (e.g., time expressions such as “2025” vs. “2024” in the second observation) from key terms (e.g., sales).
\end{challenge}

\begin{challenge}\label{sem-challenge}
\emph{Unseen information} contained in queries, which is absent from the historical data, must be handled properly. A user’s new query may resemble past ones but introduce previously unseen entities or attributes (e.g., ``Huawei'' and ``SOV'').
Such information may have to be added to the final DSL, and, in the rarer case, may need to be used for generating a new skeleton and populating the cache.

    % How can we handle novel information that is absent from historical data? A user’s new query may resemble past ones but introduce previously unseen entities or attributes (e.g., "Huawei" and "SOV"). We need to query the knowledge base and perform precise, rapid knowledge retrieval.
\end{challenge}

% Para 8 

% For \textbf{C1}, we incorporate a hybrid skeleton construction approach that includes template cache construction for historical queries and direct skeleton embeddings generation for online user queries.
We propose \model{}, a novel NL-to-DSL framework based on a query semantic cache, for business analytics. 
To address \textbf{C1}, we propose a hybrid strategy that manages skeletons in two phases: (1) \emph{offline skeleton construction}, where skeletons are first generated from historical queries via a dedicated construction pipeline and then stored in the cache (see Section~\ref{cache_construction}); and (2) \emph{online skeleton embedding}, where an entity-agnostic model pretrained via contrastive learning directly generates skeletal embeddings from user queries. These techniques provide the basis for cache creation and use (refer to Section~\ref{CL4encoder}).

% Para 9
For \textbf{C2}, we introduce a heterogeneous Retrieval-Augmented Generation (RAG) method~\cite{lewis2020retrieval} to enhance the understanding of the NL query at syntactic, data, and semantic levels. Our method integrates three specialized knowledge sources, namely (i) DSL Configuration, (ii) Column Values, and (iii) Enterprise Domain Knowledge.  This method improves the final DSL generation with the complementary knowledge provided via RAG (see Section~\ref{knowledge}).
%Each of these knowledge sources is handled by a distinct processing strategy.

% While the concise DSL configuration is directly injected into the prompt, we employ distinct retrieval strategies for the two more extensive knowledge sources. For column values, we use a hybrid retriever combining sparse (BM25) and dense embeddings, fusing the results with Reciprocal Rank Fusion (RRF). For domain knowledge, we pre-compute a Locality-Sensitive Hashing (LSH) index over its vector embeddings to enable efficient, sub-linear time retrieval.

We evaluate \model{} on six in-house datasets built from real operational data at Xiaohongshu, achieving a multifold speedup without compromising on performance. 
Furthermore, we also evaluate this system on two new NL-to-DSL benchmarks, named \newspider{} and \newbird{}, that are synthesized from the popular Text-to-SQL datasets Spider\cite{yu2018spider} and BIRD\cite{li2023can}. \model{} demonstrates substantially superior performance compared to the in-context learning (ICL) setting (see Section~\ref{section:new-datasets}).

% Para 10
To the best of our knowledge, \model{} is the first NL-to-DSL framework reported to efficiently support LLM-based business analytics. %
% Contributions
Our main contributions are summarized as follows:

\begin{itemize}
    \item Based on our industry practice at Xiaohongshu, we introduce \model{}, a novel NL-to-DSL framework that generates DSL based on historical skeleton caching, leading to \textbf{reduced inference time} and \textbf{robust performance} for enterprise-scale data.
    \item We present a hybrid skeleton construction strategy that (i) constructs a representative skeleton cache and (ii) uses skeleton-aware contrastive learning to train an entity-agnostic embedding model for accurately deriving query skeletons.
    \item We propose an RAG method to effectively augment the generation context by integrating heterogeneous knowledge from diverse sources at the syntactic, data, and semantic levels.
    % \item We publish a corporation dataset in anonymized form which consists of tens of thousands of real business operations data spanning three representative domains, namely commercial, community, and transactions. To assess the extensibility of our method, we also adapt two widely-used NL-to-SQL datasets (Spider and BIRD) to the NL-to-DSL setting. 
    
    \item \model{} achieves a substantial acceleration without compromising accuracy, a performance robustly validated on real business datasets from Xiaohongshu and new NL-to-DSL datasets adapted from widely-used Text-to-SQL datasets.

\end{itemize}

\section{Preliminary}
\label{section:preliminary}
In this section, we first outline the necessary background before providing the problem statement of NL-to-DSL in business analytics. Subsequently, we present representative data and pilot experiments to demonstrate the urgent need for query semantics caching for NL-to-DSL.

\subsection{Background}
Industry practice commonly combines denormalized wide tables with a semantic-layer DSL to deliver interactive latency, operational robustness, and consistent governance.

\textbf{Why wide tables?}
For performance and reliability, numerous industrial systems employ pre-joining and unioning in offline or near-real-time pipelines to avoid costly multi-table joins during online queries. Representative systems include (1) Facebook’s Scuba~\cite{chen2016realtime}, which drives low-latency, in-memory analytics via wide tables and pre-aggregation; (2) Google’s Mesa~\cite{gupta2016mesa}, which ensures large-scale metric consistency and low latency through pre-aggregation and materialization. A shared conclusion across these systems is that standardizing query patterns onto wide tables can significantly reduce latency and complexity. Crucially, this “wide table” is often not an original physical table in the operational schema but an aggregation or summary snapshot produced by prior cross-entity joins, thereby shifting expensive joins from the online serving path to the data production stage. At Xiaohongshu, this practice results in our vast operational data being partitioned into several dozen wide tables, each aligned with a distinct business domain.

\textbf{Why use DSL?}
Enterprises standardize metrics and security by introducing a semantic layer expressed in a constrained DSL (dimensions, metrics, filters, aggregations) that serves as a common intermediate representation. The DSL compiles to multiple backends, primarily SQL and visualization specifications. For SQL, it maps dimensions, metrics, and filters to grouping, aggregation, and WHERE/HAVING clauses, respectively, using dialect-aware rewrites. For visualizations, these same concepts are translated into visual encodings, user interactions, and summaries. Examples include Apache Calcite~\cite{begoli2018apache}, which lowers high-level expressions to relational algebra with cross-source optimization, and Zenvisage~\cite{siddiqui2021sketching}, which captures user intent in a task-oriented DSL and delegates efficient execution.

\textbf{Necessity of the long-chain approach.}
First, the prohibitive cost and complexity of annotating large-scale datasets from real user queries make data-hungry, end-to-end models impractical and non-robust.
% As shown in Fig.~\ref{fig1:dsl}, the annotation process requires dual expertise. Annotators must possess not only the technical proficiency to identify structural query components (e.g., WHERE/HAVING clauses and aggregation methods) but also deep business knowledge in choosing the appropriate tables. 
Specifically, labeling requires a simultaneous grasp of structural query components (e.g., WHERE/HAVING clauses and aggregation methods) and domain-specific business logic to prevent mapping errors.
Second, LLMs have strong few-shot capabilities that help resolve semantic ambiguity and deliver higher precision without extensive supervised training.
Third, decomposing the task into well-scoped stages provides modularity, transparency, traceability, and operational control, all of which are critical for iterative development under resource constraints.

Nevertheless, longer pipelines suffer from increased latency and computational cost, and critically, they risk amplifying errors across stages. These issues are confirmed in our preliminary analysis (see Section \ref{section:data-analysis}). To mitigate these drawbacks, we propose a caching-based shortcut that derives the target DSL from semantically similar queries stored in the cache. This approach significantly reduces both temporal and token consumption while maintaining high accuracy.

\subsection{Problem Definition}

NL-to-DSL translates an NL query into a DSL specification, a common routine in BI. DSL is a specialized, highly standardized, and structurally constrained language designed for a particular problem domain. In this context, the DSL encodes the required data and processing logic via fields: MeasureList, DimensionList, and FilterList, as defined in Table~\ref{tab:component_definitions}.

\begin{table}[!ht]
\centering
\footnotesize
\renewcommand{\arraystretch}{0.85}
\caption{Definitions of the \textbf{\textit{general}} DSL components.}
\label{tab:component_definitions}
\renewcommand{\arraystretch}{0.85}
\begin{tabular}{l m{0.7\linewidth}}
\toprule
\textbf{Component} & \multicolumn{1}{c}{\textbf{Definition}} \\
\midrule
\textbf{Measure}   & A quantitative value dynamically computed via aggregation (e.g., `SUM`, `COUNT`) or arithmetic operations on one or more fields. \\ 
\addlinespace % Adds a little extra space between rows for readability

\textbf{Dimension} & A categorical field or perspective used to group, segment, or disaggregate a measure. \\
\addlinespace

\textbf{Filter}    & A condition that constrains data by filtering rows, either before aggregation (like a `WHERE` clause) or after aggregation (like a `HAVING` clause). \\
\bottomrule
\end{tabular}
\end{table}

\textbf{Problem Statement.}
Considering a target NL query $Q$ on certain table repository $\mathcal{T}$, the target of NL-to-DSL is to maximize the possibility of generating the correct DSL $D$, formally:

\begin{definition}[NL-to-DSL]
\label{definition:nl2dsl} 
    First, the NL-to-DSL task for business analytics leverages a modeling function, $M_{prompt}$, not merely for a simple interpretation, but for complex reasoning sub-tasks such as query clarification and structural analysis. This process transforms the raw query $Q$ into an enriched query representation, $Q'$, which is then used by the retrieval function $R$ to gather the most relevant context $C$ from the table repository $\mathcal{T}$ and external knowledge $\mathcal{K}$:
    \begin{equation} \label{eq:retrieval_enhanced}
    C = R(Q', \mathcal{T}, \mathcal{K}) \quad \text{where} \quad Q' = M_{prompt}(Q)
    \end{equation}
    With the retrieved context $C$, the subsequent step is the synthesis of the final DSL specification $D$. This is not a monolithic step but a structured generation process, typically involving a sequence of model calls ($M_\mathit{gen} = M_n \circ \dots \circ M_1$) to iteratively construct different components of the DSL, such as its Measures, Dimensions, and Filters, thus named \emph{long-chain}. The overall conditional probability of generating the complete DSL $D$ is thus factorized into the probabilities of generating each component sequentially. The objective is to find the optimal DSL $D^*$ that maximizes this factorized probability:
    \begin{equation} \label{eq:generation_enhanced}
    D^* = \arg\max_{D} \prod_{i=1}^{n} P_{M_{gen_i}}(D_i | Q, C, D_{<i})
    \end{equation}
\end{definition}

To reduce the number of LLM invocations and improve efficiency, we propose a \emph{two-step shortcut approach}: (1) first to retrieve skeleton that exhibit structural similarity from template cache; (2) second to synthesize the target DSL using the retrieved DSLs as in-context exemplars.

Specifically, the shortcut NL-to-DSL path encodes the user query $Q$ into its corresponding skeletal embedding $S=E(Q)$, using a pre-trained, entity-agnostic model $E$. Given the established query semantics cache $\mathcal{C} =\{ (S'_i, D'_i) \}_{i=1}^n$, the optimal DSL $D^*$ equation can be reformulated as:
\begin{equation} \label{eq:generation_enhanced}
D^* = \arg\max_{D} \prod_{i=1}^{n} P_{M_\mathit{gen}}(D_i \mid S, R, \mathcal{K}, D_{<i})
\end{equation}
where $R = \{ D'_i \mid i \in \text{topK}_{j=1 \dots n}(\sigma(S, S'_j)) \}$ represents the set of retrieved DSLs, $\sigma$ is a relevance scoring function, and $(S'_j, D'_j) \in \mathcal{C}$. The entire process involves only a single LLM call $M_\mathit{gen}$.

\subsection{Data Analysis}
\label{section:data-analysis}
In this section, we demonstrate the necessity and feasibility of the shortcut NL-to-DSL approach based on skeleton caching by analyzing its resource efficiency and data distribution.

\begin{table}[ht]
\centering
\footnotesize
\caption{Cost statistics of the long-chain pipeline}
\label{pre-cost}
\renewcommand{\arraystretch}{0.85}
\begin{tabular}{cccc}
\toprule
\textbf{Steps} & \textbf{P90~(s)} & \textbf{\#(Token)} & \textbf{Accuracy} \\
% \hhline{-|-|-|-}
\midrule
Problem Analysis            & 6.00  & 8,821  & 94.07\\

Data Retrieval              & 5.05  & 4,043  & 66.95\\

Dimension Generation        & 4.60  & 3,901  & 56.67\\

% Measure Generation          & 7.20  & 3,493  & 60.00\\
Measure Generation          & 7.20  & 3,741  & 60.00\\

Filter Generation           & 7.40  & 5,765  & 36.67\\
% \hhline{-|-|-|-}
\midrule
% Total                       & 30.25 & 26,023 & 35.00\\
Total                       & 30.25 & 26,271 & 35.00\\
\bottomrule
\end{tabular}
\label{tab:data-analysis}
\end{table}

\subsubsection{Cost Analysis}
\label{2-Cost-Analysis}
% We conducted a step-by-step analysis of the long-chain process. Fig.~\ref{fig2:main} illustrates the stages of this pipeline, which typically include at least problem understanding, data retrieval, and data analysis. The goal of problem understanding is to grasp the context of the query and better infer the user’s intent; When necessary, this involves operations such as keyword extraction or expansion, typically by invoking LLM APIs. The goal of data retrieval is to fetch data relevant to the query from enterprise-scale data lakes, commonly using BM25\cite{robertson2009probabilistic} or vector search\cite{platzer2005vector}. The goal of data analysis is to interpret the retrieved data and formulate a DSL capable of answering the user’s query; this is usually achieved by prompting an LLM with the user’s query together with the retrieved data.
We analyzed the long-chain pipeline (Fig.~\ref{fig2:main}), comprising three stages: (1) \textbf{Problem Analysis} for clarification and intent inference via LLM APIs; (2) \textbf{Data Retrieval} from enterprise data lakes using BM25~\cite{robertson2009probabilistic} or vector search~\cite{platzer2005vector}; and (3) \textbf{DSL Generation}, where an LLM formulates the target DSL using the retrieved context.

We report the average cost of each step computed over the community dataset, as shown in Table~\ref{pre-cost}. Due to multiple invocations of LLMs, the total latency~(measured by P90, which is formally defined in~\ref{section:metrics}) of 30.25 seconds significantly exceeds the acceptable threshold for real-time user response. Furthermore, excessive token length (a total of 26,023 tokens) results in substantial resource waste. Ultimately, the multi-step nature leads to error propagation, resulting in a final accuracy of only 35.00\%, which is far from ideal. Therefore, it is imperative to reduce both the number of processing steps and the frequency of LLM calls.

\begin{figure}[htbp]
    \centering
    \begin{subfigure}[t]{0.3\columnwidth}
        \centering
        \includegraphics[width=1\textwidth]{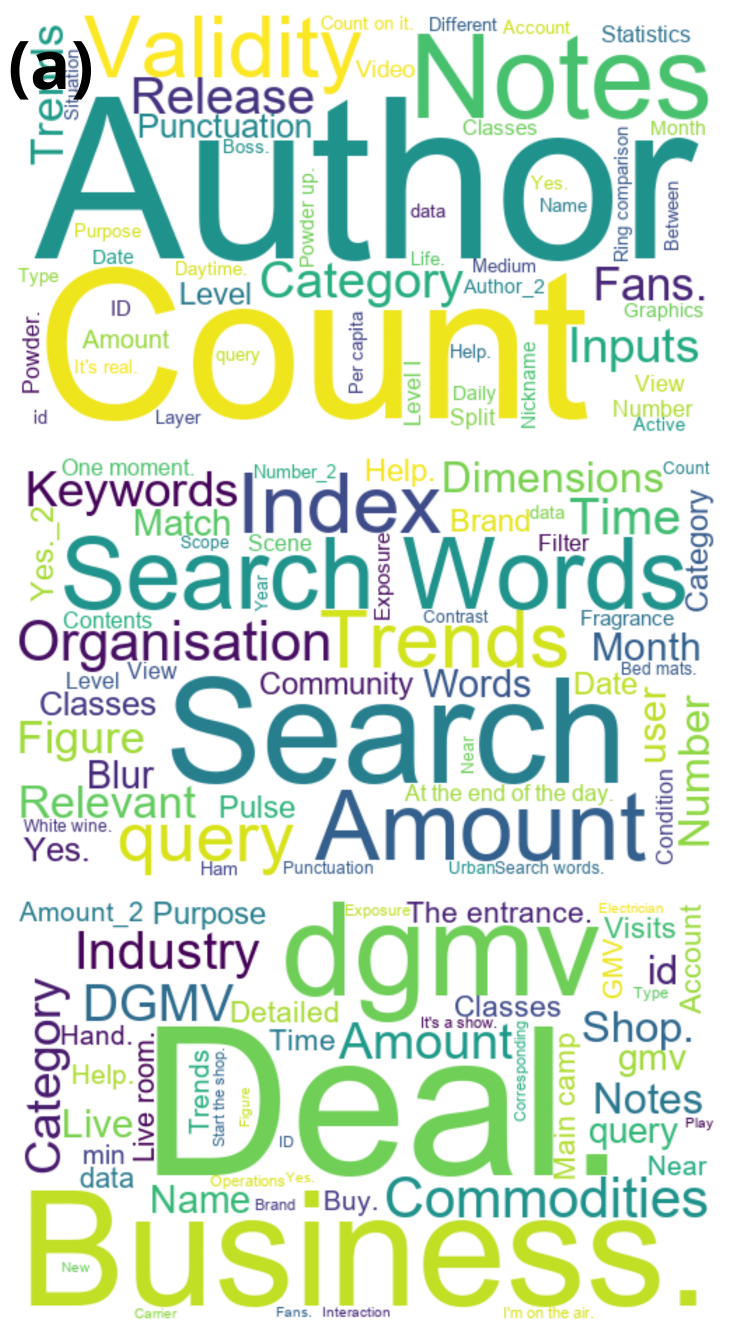}
        % \caption{Word cloud of query skeletons.}
        \label{fig3:wordcloud-1}
    \end{subfigure}%
    ~ 
    \begin{subfigure}[t]{0.635\columnwidth}
        \centering
        \includegraphics[width=1\textwidth]{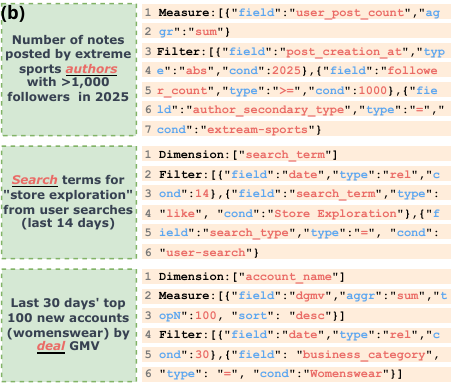}
        % \caption{Examples of queries and their corresponding DSLs.}
        \label{fig3:wordcloud-2}
    \end{subfigure}
    \caption{(a) Visualization of query skeleton clusters and (b) corresponding examples of user queries and their DSLs. Skeletons within the same cluster share similar structural patterns in their DSL representations.}
    \label{fig3:wordcloud}
    % 作者、数量
    % 搜索，搜索词
    % 成交、商家
\end{figure}

\subsubsection{Data Statistics}
We conducted a skeleton cluster analysis on the enterprise dataset. As shown in the Fig.~\ref{fig3:wordcloud}, the key findings are as follows:
\begin{itemize}
    \item The skeleton of user queries exhibit significant similarity. From Fig.~\ref{fig3:wordcloud}~(a), it can be observed that the similarity within clusters is extremely high, often sharing reusable keywords with the skeleton.
    
    \item As demonstrated in Fig.~\ref{fig3:wordcloud}~(b), similar queries tend to generate similar DSLs. More specifically, queries with similar skeletons generally correspond to DSLs with high resemblance.
\end{itemize}

\remarkbox{
The traditional long-chain NL-to-DSL pipeline incurs substantial latency, token consumption, and exacerbates error propagation; Moreover, optimizations applied within this pipeline yield only limited gains in accuracy and modest reductions in latency. Consequently, developing more concise pipelines is imperative. Our data analysis shows that the skeletons of user queries exhibit high similarity, and similar skeletons typically map to similar DSLs, making a shortcut approach—leveraging the DSLs of similar queries to generate the target DSL—highly feasible.
}
\begin{figure*}[htbp]
\centerline{\includegraphics[width=1.65\columnwidth]{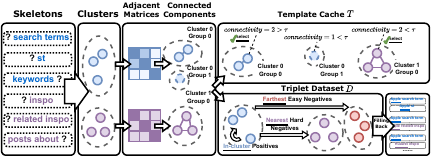}}
\caption{An overview of the construction process for the skeleton cache (top right) and the triplet dataset (bottom right). Both processes share a common preprocessing pipeline: (1) query skeletons are first grouped into coarse clusters. (2) Then, within each cluster, skeletons are further partitioned into fine-grained groups by identifying the connected components in a similarity graph, which is constructed by thresholding the pairwise adjacency matrices.}
\label{fig:data-preparation}
\end{figure*}

\section{Overview}

This section presents an overview of our proposed NL-to-DSL framework \model{}. The core design of \model{} is a shortcut solution for NL-to-DSL, as illustrated in Fig.~\ref{fig2:main}. 
% Additionally, if the skeleton cache fails to yield satisfactory results, the framework falls back to a failsafe long-chain pipeline, though this is not detailed in the paper.
In our production environment, in case of a skeleton cache miss (i.e., no high-similarity match is found), the framework defaults to a long-chain pipeline.

Our proposed solution works in offline and online phases. 
The offline phase involves the following two steps:
\begin{itemize}
    \item \textbf{Skeleton Cache Construction.}
    We first construct query skeletons via ICL, which are refined using a pipeline of named entity recognition (NER)~\cite{HanLP}, rule-based filtering, and manual verification. Then, we apply clustering and graph-based classification to this refined set to derive a canonical collection of representative skeletons. Finally, the template cache is constructed from pairs of skeletons and their associated DSLs.
    
    \item \textbf{Entity-agnostic Embedding Model Training.}
    We propose a skeleton-based, self-supervised contrastive learning method to fine-tune the embedding model. This design allows the encoder to capture structural similarity directly, circumventing the conventional two-step workflow of explicit skeleton extraction followed by retrieval. This integrated approach improves efficiency and mitigates the error propagation inherent in multi-stage systems.
\end{itemize}

During the online phase, our approach incurs only one LLM invocation for every user request. 
%by retrieving queries similar to the input and using them as in-context demonstrations to generate the target DSL. 
% online workflow
Given a user-submitted NL query, we first perform vector search with a pretrained embedding model, and then conduct knowledge retrieval based on the query and the retrieved data. 
Subsequently, we feed the query, the retrieved data, and the retrieved knowledge into a rewriting LLM to produce the target DSL.
\begin{itemize}
    \item \textbf{(Skeleton, DSL) Pair Retrieval.} At inference time, the trained embedding model first computes a skeletal embedding for the incoming user query. This embedding is then used to perform a nearest-neighbor search against the template cache, retrieving the most similar (skeleton, DSL) pair. Finally, this retrieved pair serves as a one-shot example for the subsequent LLM-based rewriting stage.
    
    \item \textbf{Knowledge Retrieval.} 
    To address the orthogonal syntactic, data, and semantic challenges of NL-to-DSL translation, we introduce a multi-source RAG framework that integrates three specialized knowledge sources: DSL Configuration Knowledge, Column Value Knowledge, and Enterprise Domain Knowledge. While the concise DSL knowledge is injected directly into the prompt, we design specialized and efficient retrieval mechanisms for the vast Column Value and Enterprise Domain knowledge.
    
    \item \textbf{DSL Rewriting.} This module leverages the powerful contextual understanding capabilities of LLMs to generate the target DSL. We use the retrieved (Skeleton, DSL) pairs as examples and the retrieved domain knowledge as supplementary information, forming a structured instruction for the LLM to produce the final DSL.

\end{itemize}

The next section shall describe the detailed techniques employed during the offline and online phases. 
\section{Methodology}
% In this section, we describe the offline phase of RedParrot, which comprises two key components: skeleton cache construction and entity-agnostic embedding model training.
% % 
% Next, we present a skeleton-based contrastive learning approach that trains an entity-agnostic embedding model in a self-supervised fashion. We then present the Knowledge Incorporation module of RedParrot and detail an LLM-based DSL generation method that produces DSLs by leveraging the recalled (skeleton, DSL) pairs together with the incorporated domain knowledge.

% \subsection{Skeleton-based Retrieval}
% Skeleton-based Retrieval consists of two main parts: offline construction of the template cache, and online skeleton retrieval (see Figure~\ref{fig:data-preparation}).

\subsection{Skeleton Cache Construction}
\label{cache_construction}
The core of the shortcut NL-to-DSL approach lies in bypassing the multi-stage pipeline by directly generating the DSL for a user's query from matched (skeleton, DSL) pairs in the historical cache.
However, using all historical data as the skeleton cache is impractical because there are a few special cases, and for efficiency the historical cache should not be excessive in size.
Therefore, we conclude that the skeleton cache should be compact, high-quality, and periodically refined through iterative updates.
Accordingly, our skeleton cache construction includes: (1) skeleton extraction to remove distractor words and improve generality; (2) clustering to preliminarily aggregate similar skeletons; (3) building a connectivity graph to obtain a more fine-grained measure of skeleton similarity; and (4) ultimately creating a skeleton cache of skeletons and their corresponding DSLs.

\paragraph{Skeleton Extraction} As illustrated in Example~\ref{ex2}, queries can be semantically similar due to shared distractor words yet map to different DSLs because they reference distinct tables. The primary objective of skeleton extraction is thus to eliminate these semantic distractors. To this end, we employ a hybrid approach. 
Initially, we leverage an LLM's inference capability by providing it with several high-fidelity query-skeleton pairs derived from targeted manual corrections of previously erroneous DSL samples. This enrichment step, while optional, ensures the fundamental accuracy of the underlying DSLs, which in turn facilitates more precise skeleton construction.
% revised annotation
% \change{Initially, we leverage an LLM's inference capability by providing it with several high-fidelity query-skeleton pairs derived from targeted manual\rone{R1.C3} corrections of previously erroneous DSL samples. This enrichment step, while optional, ensures the fundamental accuracy of the underlying DSLs, which in turn facilitates more precise skeleton construction.}
% 
This output is then refined through a two-stage post-processing procedure: first, we apply entity recognition to excise any remaining distractors, and second, we use rule-based regular-expression transformations to normalize its structure and ensure formatting correctness.

\paragraph{Clustering} The goal is to perform a preliminary classification of the generated skeletons based on similarity. We first use an encoder to map the skeletons into a vector space (specifically, we employ qwen3-embedding-4b), and then apply k-means~\cite{ahmed2020k} clustering in this vector space to obtain $M$ clusters.

\paragraph{Connectivity Graph Construction} The goal is to perform finer-grained intra-cluster similarity computation on the previous clustering results and to select representative skeletons to add to the skeleton cache. Specifically, we treat each (skeleton, DSL) data point as a vertex, and use the similarity between queries as edges. Each connected component is taken as a group, and vertices within a group whose degree is greater than 4 are selected for inclusion in the skeleton cache.

% \paragraph{Connectivity Graph Construction} The goal is to perform finer-grained intra-cluster similarity computation on the previous clustering results and to select representative skeletons to add to the skeleton cache, as shown in Algorithm~\ref{template-selection}. We treat each (skeleton, DSL) data point as a vertex, and use the similarity between queries as edges. Each connected component is taken as a group, and vertices within a group whose degree is greater than 4 are selected for inclusion in the skeleton cache.

% In the real-world deployment of our shortcut NL-to-DSL pipeline, when the number of new queries reaches a predefined threshold parameter, the skeleton cache undergoes a full rebuild. This process involves reapplying the algorithms described above to the complete set of historical and recent skeletons, which ensures the cache's structural patterns reflect comprehensive insights.

During the system's cold-start phase or when encountering novel queries that fall below the predefined semantic similarity threshold, \model{} defaults to the long-chain generation pipeline. The resulting verified query-DSL pairs are then leveraged to bootstrap the semantic cache and facilitate continuous incremental population. To counter metadata drift, a heterogeneous RAG module leverages RRF-based hybrid retrieval to synchronize cached skeletons with the latest schema and data representations. This mechanism ensures that even for skeletons retrieved from the cache, the final DSL reflects the most current schema state.
Furthermore, the system maintains cache vitality through two distinct update strategies: a periodic global rebuild for comprehensive consistency and an online incremental update for efficiency. More details about the cache updating are discussed in Section~\ref{sec:cache_update}.

% \change{\rone{R1.W2}During the system's cold-start phase or when encountering \rtwo{R2.W3}novel queries that fall below the predefined semantic similarity threshold, \model{} defaults to the long-chain generation pipeline. The resulting verified query-DSL pairs are then leveraged to bootstrap the semantic cache and facilitate continuous incremental population. To counter metadata drift, a heterogeneous RAG module leverages RRF-based hybrid retrieval to synchronize cached skeletons with the latest schema and data representations. This mechanism ensures that even for skeletons retrieved from the cache, the final DSL reflects the most current schema state.}
% % 
% \change{\rthree{R1.C6, R2.C2, R3.W1} Furthermore, the system maintains cache vitality through two distinct update strategies: a periodic global rebuild for comprehensive consistency and an online incremental update for efficiency. More details about the cache updating are discussed in Section~\ref{sec:cache_update}.}

\subsection{Entity-agnostic Embedding Model}
\label{CL4encoder}
In this section, we present \model{}'s skeleton-based contrastive learning module~\cite{khosla2020supervised,mai2024learning}, which constructs positive and negative pairs using skeleton similarity to fine-tune a PLM-based encoder.

The aim is to fine-tune a pre-trained language model (PLM)-based encoder so that it can recognize similarity in the skeleton while reducing the influence of distracting words. We adopt a contrastive learning approach that automatically constructs skeleton-based positive and negative pairs to train the PLM-based encoder in an unsupervised manner.

Contrastive learning is a self-supervised approach that shapes representations so that similar inputs are proximal and dissimilar inputs are well separated. Our objective is to learn an encoder $\theta$ (e.g., a query encoder) that maps user queries to a high-dimensional vector. By pulling together embeddings of queries that share similar skeletons and pushing apart those with dissimilar ones, we aim to endow the encoder with the ability to strengthen the skeleton signal while retaining sufficient contextual information. The most straightforward approach would be to add an additional step online that extracts the skeleton and removes distractor tokens, followed by skeleton-based retrieval. However, performing skeleton extraction at inference time introduces an extra stage that increases latency and propagates errors; Moreover, the removed segments may carry some contextual information. Strengthening the skeleton while retaining sufficient contextual information through a fine-tuned encoder is therefore a faster and more accurate approach.

To fine-tune the encoder, we generate training pairs by partitioning samples through skeleton extraction, clustering, and connectivity graph construction. For a given anchor selected from a connected component, we construct a triplet: positive samples are drawn from the same component, hard negatives from the same cluster but different components, and negatives from disjoint clusters. To leverage the triplet, we apply a contrastive loss that minimizes the distance between skeleton-similar queries while maximizing it for distinct ones. The loss $\mathcal{L}_{\text{pair}}(i,j)$ for a query pair $(q_i, q_j)$ is first defined as:

\begin{equation}
    \mathcal{L}_{\text{pair}}(i,j) = (-1)^{\mathbf{1}_{[\{q_i,q_j\}\subseteq g]}}\lVert q_i - q_j \rVert_2
\end{equation}

where $\mathbf{1}_{[\{q_i,q_j\}\subseteq g]}$ is a sign function equal to 1 when $q_i$ and $q_j$ are from the same connected components, i.e., they are in a matched pair. Next, we can obtain the contrastive loss by averaging over all matched and unmatched pairs, where $m$ is the margin and $\alpha$ is a weighting hyperparameter:

\begin{equation}
    \mathcal{L}(\theta) = \frac{1}{n}\sum_{(q_i, q_j, q_j)\in D}[\alpha\mathcal{L}_{\text{pair}}(i,j) + (1-\alpha)\mathcal{L}_{\text{pair}}(i,k) + m]_+
\end{equation}

\subsection{Skeleton Retrieval}
This step retrieves relevant (skeleton, DSL) pairs for user query $Q$ from skeleton cache $T$. Using a fine-tuned encoder $\theta$ (see Section~\ref{CL4encoder} for more details), all representative queries are encoded into skeletal vectors and stored in a vector database V. Given user query $Q$, \model{} encodes it as $\theta(t)$ and performs a similarity search to retrieve the top-K most similar tuples.

These retrieved (skeleton, DSL) pairs serve two purposes: (1) they provide demonstrations that guide the LLM to generate the final DSL, and (2) they identify the concrete table in the database on which the DSL operates. Resolving the correct table name is crucial: even if the DSL remains the same, using an incorrect table name will lead to incorrect execution results. We therefore select the table via weighted voting:

\begin{equation}
    TableID^* = \arg\max_{t \in T} \sum_{i=1}^{K} similarity(\widetilde{q_i},\widetilde{q_j}) \cdot \mathbf{1}_{[t_i = t]},
\end{equation}
where $similarity(q, q_i)$ denotes the similarity between the i-th retrieved query and the input query. $\mathbf{1}_{[t_i = t]}$ denotes the indicator function, which equals 1 when $t_i = t$ and 0 otherwise.

\subsection{Knowledge Construction}
\label{knowledge}

In this section, we detail the construction of multi-source knowledge RAG framework, which draws upon three principal knowledge sources: DSL Configuration Knowledge, Column Value Knowledge, and Enterprise Domain Knowledge.

\subsubsection{Multi-Source Knowledge Components}
Each of these components plays a distinct and complementary role in the query translation pipeline. They are responsible for ensuring DSL configuration correctness, performing column value linking and providing semantic context.

\paragraph{DSL Configuration Knowledge}
This knowledge source is a static set of rules defining the syntax, valid parameters, and data type constraints (e.g., \texttt{string}, \texttt{int}, \texttt{date}) of our target DSL. For example, a rule for the \texttt{string} data type maps NL predicates to logical operators: expressions like ``equals'' or ``='' are mapped to an exact match, while expressions such as ``contains'' or ``is about'' are mapped to a case-insensitive substring match.

\paragraph{Column Value Knowledge}
The knowledge source is responsible for mapping entity values expressed in NL to their canonical representations in the database. This semantic mapping process is essential for reconciling varied linguistic expressions into a unified, queryable format, thereby ensuring data consistency and integrity. For instance, consider an entity such as \texttt{Primary Product Line}. The knowledge base defines its set of enumerated canonical values (e.g., \texttt{Brand Ads}, \texttt{Performance Ads}). Crucially, it also specifies mappings for their recognized aliases; NL expressions like ``bidding'' or ``auction ads'' are both resolved to the single canonical value \texttt{Performance Ads}.

\paragraph{Enterprise Domain Knowledge}
The domain knowledge supplies business-specific semantics by maintaining a repository of specialized terminology, acronyms, and idiomatic business definitions. For example, in an e-commerce context, a user might ask, ``What is the DGMV for iPhone 17?''. The acronym ``DGMV'' stands for ``Direct Gross Merchandise Volume,'' a highly specialized business term. 
However, in the underlying database, this metric might correspond to columns with non-obvious names, such as \texttt{direct\_gmv} or \texttt{transaction\_amount\_direct}. 
Without domain knowledge to resolve this mapping, a model cannot comprehend the query's true semantics and thus fails to translate it into the correct database query.

\subsubsection{Knowledge Retrieval}
Given the distinct characteristics of these knowledge sources, we employ different methods to retrieve and integrate them into the final context. Whereas the compact DSL knowledge can be fully injected into the prompt, the vast repositories of the other two knowledge types require a sophisticated retrieval mechanism to source the relevant, query-specific context.

The retrieval process for the column value knowledge begins by isolating the specific column value from the user's query. We achieve this by computing a set difference between the original query and its corresponding skeleton query. The extracted value, denoted as the query $q$, is processed by a hybrid retrieval pipeline to identify the optimal column value knowledge. First, we retrieve the top-$k$ candidates from two complementary ranking systems. A sparse retriever, utilizing BM25, produces a lexically-ranked list $L_s = \{ (c_{s,i}, i) \}_{i=1}^{k}$, where $c_{s,i}$ is the $i$-th candidate. Concurrently, $q$ is encoded into a dense vector $\bm{v}_q = \text{Encode}(q)$, which is used to query a semantic index (e.g., FAISS) to yield a semantically-ranked list $L_d = \{ (c_{d,i}, i) \}_{i=1}^{k}$.

The ranked lists $L_s$ and $L_d$ are subsequently aggregated using RRF. For each unique candidate $c$ in the union of candidates from both lists, denoted by $C$, we compute a fused score $S(c)$ as follows:
\begin{equation}
\label{eq:rrf}
S(c) = \frac{1}{k_{rrf} + \operatorname{rank}_{L_s}(c)} + \frac{1}{k_{rrf} + \operatorname{rank}_{L_d}(c)}
\end{equation}
where $\operatorname{rank}_L(c)$ is the rank of candidate $c$ in list $L$ (defined as $\infty$ if $c$ is not in $L$), and $k_{rrf}$ is a constant that mitigates the impact of high-ranking items. The final mapping, $m^*$, is then selected as the candidate with the highest fused score:
\begin{equation}
\label{eq:argmax}
m^* = \underset{c \in C}{\operatorname{argmax}} \, S(c)
\end{equation}
This approach robustly combines signals from both lexical and semantic spaces without requiring score normalization, yielding a single, high-confidence mapping.

% To efficiently ground user queries in our large and evolving domain-specific knowledge base, we leverage LSH for sub-linear time approximate nearest neighbor (ANN) search. In an offline preprocessing step, we first generate dense vector embeddings for all term-definition pairs in our lexicon using a pre-trained language model. These embeddings are then indexed into an LSH data structure, where multiple hash functions map semantically similar vectors to the same buckets with high probability. At runtime, candidate spans from the input query are embedded into the same vector space. We then probe the LSH index with the query embedding to retrieve a small set of candidate definitions from the colliding buckets. 

To ensure efficient grounding in our evolving knowledge base, we employ LSH for sub-linear ANN search. Offline, term-definition pairs are embedded and indexed into LSH buckets based on semantic similarity. At runtime, query embeddings probe the index to retrieve candidates from colliding buckets. This rapid retrieval mechanism allows our system to dynamically resolve semantic ambiguities, ensuring the final synthesized DSL representation accurately captures the user's business intent.

\begin{figure}[htbp]
\centerline{\includegraphics[width=0.8\columnwidth]{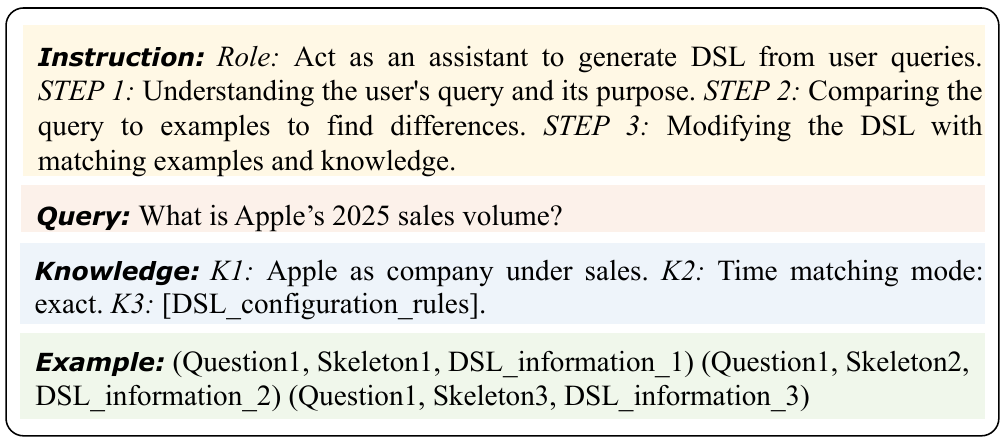}}
\caption{An example of a DSL rewriting prompt.}
\label{fig:embedding-model}
\end{figure}

\subsection{DSL Rewriting}
In this section, we present \model{}'s LLM-based DSL generation module that leverages retrieved (skeleton, DSL) pairs and domain knowledge, operating in a training-free, few-shot setting.

Our earlier high-precision retrieval of (skeleton, DSL) pairs, along with the supplementation of additional knowledge, has laid a solid foundation for this training-free, few-shot setting. Here, we use Qwen-2.5-Instruct together with a carefully designed prompt. The prompt primarily includes: (1) (skeleton, DSL) pairs; and (2) retrieved business/domain-specific knowledge, as shown in Fig.~\ref{fig:embedding-model}.

\section{Experiments}
In this section, we present a comprehensive experimental evaluation of \model{}. Our experiments are conducted on a combination of real-world enterprise datasets and our newly adapted NL-to-DSL versions of the widely-used Text-to-SQL benchmarks, Spider~\cite{yu2018spider} and BIRD~\cite{li2023can}, which we refer to as \newspider{} and \newbird{}. The evaluation is designed to answer the following key research questions:
\begin{itemize}
    \item \textbf{RQ1}. How effectively does \model{} accelerate the NL-to-DSL process while maintaining accuracy?
    % 主实验
    % 我们框架的表现是否符合我们的预期？加速且保持准确率
    \item \textbf{RQ2}. How well does \model{} generalize to non-enterprise, general-domain data?
    % Spider/Bird
    % 我们的任务/方法的通用性如何
    \item \textbf{RQ3}. What is the contribution of each proposed technique in \model{} to its overall performance?
    % Ablation Study
    % 每个小技术点的有效性
    \item \textbf{RQ4}. How effective is the proposed entity-agnostic embedding model at retrieving structurally similar templates for a given query?
    % Visualization, Skeleton & Table Settings
    % 我们提出的方法何种程度上解决了我们的问题
    \item \textbf{RQ5}. How can the semantic cache be effectively updated?
    
\end{itemize}

% 实验：
% 1. 主实验：短链路在可以进短链路的数据（百来条）上和长链路的准确率（每个步骤的准确率）、overheads（时间、token cost）
% 2. 不擦、ner简单擦、人工擦、entity-agnostic embedding最后改写后准确率的比较
% 2.1 entity-agnostic embedding和人工擦的聚类图
% 3. abltion：knowledge incorporation

\subsection{Experimental Settings}

\subsubsection{Datasets}
Our experimental evaluation is designed to assess the performance of \model{} in two primary contexts: a real-world enterprise environment and general open-source benchmarks. For the enterprise evaluation, we curated three datasets from internal data provided by Xiaohongshu, each corresponding to a core business domain: \commerce{}, \community{}, and \trading{}. According to the implementations discussed in Section~\ref{section:implemantations}, two versions of each dataset were prepared, yielding a total of six experimental settings. A statistical summary of these datasets is presented in Table~\ref{tab:datasets}. To evaluate the model's generalization capabilities, we also constructed two new open-source NL-to-DSL datasets \newspider{} and \newbird{}, by adapting the popular Spider and Bird benchmarks. The creation process and details of these datasets are described in Section~\ref{section:new-datasets}.

\begin{table}[htbp]
\footnotesize
\renewcommand{\arraystretch}{0.75} % 稍微增加行高以提升可读性
\caption{Dataset statistics}
\label{tab:datasets}
\begin{center}
{
    \begin{tabular}{@{}l cc cc@{}} 
    \toprule
    \multirow{2}{*}{\textbf{Dataset}} & \multicolumn{2}{c}{\textbf{\texttt{-095}}} & \multicolumn{2}{c}{\textbf{\texttt{-0916}}} \\
    \cmidrule(lr){2-3} \cmidrule(l){4-5} % 添加横线区分两组数据
     & \textbf{\#~Test} & \textbf{\#~Train} & \textbf{\#~Test} & \textbf{\#~Train} \\
    \midrule
    \textbf{\texttt{RED-commerce}}   & 445 & \multirow{3}{*}{4494} & 373 & \multirow{3}{*}{4310} \\
    \textbf{\texttt{RED-community}}  & 66  &  & 61  &  \\
    \textbf{\texttt{RED-trading}}    & 31  &  & 24  &  \\
    \bottomrule
    \end{tabular}
}
\end{center}
\end{table}

% \begin{table}[htbp]
% \footnotesize
% \renewcommand{\arraystretch}{0.8}
% \caption{Dataset statistics.}
% \begin{center}
% {
%     \begin{tabular}{@{}c | c | c@{}} 
%     \toprule
%     \textbf{Dataset} & \textbf{\#~Test} &   \textbf{\#~Train} \\
%     \midrule
%     \commerceCoveredL{}    &   445     &     \multirow{3}{*}{4494} \\
%     \communityCoveredL{}   &   66     & \\
%     \tradingCoveredL{}     &   31      & \\
%     \midrule
%     \commerceCovered{}    &   373     &     \multirow{3}{*}{4310} \\
%     \communityCovered{}   &   61      & \\
%     \tradingCovered{}     &   24      & \\
%     \bottomrule
%     \end{tabular}
% }
% \label{tab:datasets}
% \end{center}
% \end{table}

\begin{table*}[htbp]
    \caption{Main results.}
    \label{tab:main-a-final} 
    \centering

    \renewcommand{\arraystretch}{0.85} % 压缩行高，使表格更紧凑

    % \begin{adjustbox}{width=\textwidth}
        
        % \parbox{\textwidth}{
        \centering 
        % --- 第一个表格 ---
        {
        \begin{tabular}{@{}c | cccccc | cccccc@{}} 
        \toprule
        
        \multirow{2}{*}{\textbf{Methods}}   & \multicolumn{6}{c|}{\textbf{RED-Commerce-095}} & \multicolumn{6}{c}{\textbf{RED-Commerce-0916}} \\
        
        & TB     & DM     & MS   & FT    & \textbf{ACC}  & \textbf{P90}   & TB    & DM   & MS   & FT   & \textbf{ACC}  & \textbf{P90} \\

        \midrule

        \longchainDS{}      & 69.66	                & 82.7              & 53.93             & 97.53	            & \textbf{45.62}	            & \textbf{22.00}	            & 70.25	            & 80.97	    & 51.48	    & 95.98	    & \textbf{45.84}	    & \textbf{25.00}    \\

        \longchainQwen{}      & 15.05	                & 66.42	            & 47.19	            & 92.81	            & \textbf{9.66}	            & \textbf{21.00}	            & 18.5              & 64.76	    & 47.19	    & 90.62	    & \textbf{11.53}	    & \textbf{22.00}    \\

        \midrule

        % \modelRaw{}   & 77.93    & 72.52	 & 66.89	 & 83.56	 & 44.37 (\textbf{$\downarrow$1.25})	  & 7.16 (\textbf{$\times$3.33})                      & 87.67	   & 78.02	 & 65.95	 & 86.06	 & 47.45 (\textbf{$\uparrow$1.61})	  & 6.63 (\textbf{$\times$4.87})  \\

        \modelRaw{}   & 77.93    & 72.52	 & 66.89	 & 83.56	 & \textbf{44.37}	  & \textbf{7.16}                      & 87.67	   & 78.02	 & 65.95	 & 86.06	 & \textbf{47.45} 	  & \textbf{6.63}  \\
        
        % \modelMan{}   & 86.94	 & 79.28     & 70.27	 & 96.85	 & 51.13 (\textbf{$\uparrow$5.51})	  & 10.43 (\textbf{$\times$2.43})	            & 91.15	            & 84.99	    & 72.39	    & 93.57	  & 58.45 (\textbf{$\uparrow$12.61})	    & 10.45 (\textbf{$\times$3.05}) \\

        \modelMan{}   & 86.94	 & 79.28     & 70.27	 & 96.85	 & \textbf{51.13} 	  & \textbf{10.43} 	            & 91.15	            & 84.99	    & 72.39	    & 93.57	  & \textbf{58.45}	    & \textbf{10.45}  \\

        % \model{}       & 85.36	  & 76.13	& 69.37	     & 91.22	 & 48.65 (\textbf{$\uparrow$3.03})	   & 7.16 (\textbf{$\times$3.33})	            & 90.62	            & 81.5	    &70.24	    & 91.42	    & 57.64 (\textbf{$\uparrow$11.80})	    & 7.15 (\textbf{$\times$4.39})  \\

        \model{}       & 85.36	  & 76.13	& 69.37	     & 91.22	 & \textbf{48.65} 	   & \textbf{7.16} 	            & 90.62	            & 81.5	    &70.24	    & 91.42	    & \textbf{57.64}	    & \textbf{7.15}   \\

        \rowcolor{blue!10}
        $\Delta$ &+15.7    &-6.57     &+15.44     &-6.31      &\textbf{+3.03}      &\textbf{$\times$3.33}       &+20.37     &+0.53      &+18.76     &-4.56      &\textbf{+11.8}      &\textbf{$\times$4.39}       \\

        \bottomrule
        \end{tabular}
        }

        \vspace{2pt} % 表格间的垂直间距

        \centering 
        % --- 第二个表格 ---
        {
        \begin{tabular}{@{}c | cccccc | cccccc@{}} 
        \toprule
        \multirow{2}{*}{\textbf{Methods}} & \multicolumn{6}{c|}{\textbf{RED-Community-095}} & \multicolumn{6}{c}{\textbf{RED-Community-0916}} \\
        
        & TB & DM & MS & FT & ACC & P90 & TB & DM & MS & FT & ACC & P90 \\
        
        \midrule

        \longchainDS{}          & 68.18    & 68.18    & 59.09    & 89.39    & \textbf{59.09}    & \textbf{26.00}       & 68.85     & 68.85    & 60.65    & 95.08    & \textbf{60.67}    & \textbf{34.00}   \\
        \longchainQwen{}          & 51.52    & 68.18    & 57.57    & 92.42    & \textbf{43.94}    & \textbf{30.00}       & 49.18     & 63.93    & 59.02    & 95.08    & \textbf{47.54}    & \textbf{25.00}    \\

        \midrule
        
        % \modelRaw{}             & 90.91                  & 90.91                  & 66.67                  & 83.33                  & 53.03 (\textbf{$\downarrow$6.06})                  & 7.8 (\textbf{$\times$3.33})                  & 86.89                  & 77.05                  & 72.13                  & 86.89                  & 52.46 (\textbf{$\downarrow$8.21})                  & 6.98 (\textbf{$\times$4.87})                 \\
        
        \modelRaw{}             & 90.91                  & 90.91                  & 66.67                  & 83.33                  & \textbf{53.03}                  & \textbf{7.8}                   & 86.89                  & 77.05                  & 72.13                  & 86.89                  & \textbf{52.46}                  & \textbf{6.98}                  \\
        
        % \modelMan{}             & 89.39                  & 98.48                  & 78.79                  & 95.45                  & 65.15 (\textbf{$\uparrow$6.06})                 & 10.66 (\textbf{$\times$2.43})                & 88.52                  & 81.97                  & 73.77                  & 80.33                  & 54.1 (\textbf{$\downarrow$6.57})                   & 11.12 (\textbf{$\times$3.05})                \\

        \modelMan{}             & 89.39                  & 98.48                  & 78.79                  & 95.45                  & \textbf{65.15}                 & \textbf{10.66}                 & 88.52                  & 81.97                  & 73.77                  & 80.33                  & \textbf{54.1}                   & \textbf{11.12}                 \\

        % \model{}                 & 90.91                  & 98.48                  & 78.79                  & 95.45                  & 66.67 (\textbf{$\uparrow$7.58})                 & 7.8 (\textbf{$\times$3.33})                 & 90.16                  & 81.97                  & 77.05                  & 86.89                  & 59.02 (\textbf{$\downarrow$1.65})                  & 7.74 (\textbf{$\times$4.39})                 \\

        \model{}                 & 90.91                  & 98.48                  & 78.79                  & 95.45                  & \textbf{66.67}                 & \textbf{7.8}                 & 90.16                  & 81.97                  & 77.05                  & 86.89                  & \textbf{59.02}                  & \textbf{7.74}                  \\

        \rowcolor{blue!10}
        $\Delta$    &+22.73  & +30.3  & +19.7  & +6.06  & \textbf{+7.58}  & \textbf{$\times$3.33}  & +21.31  & +13.12  & +16.4  &  -8.19  &  \textbf{-1.65}  & \textbf{$\times$4.39}  \\
        
        \bottomrule
        
        \end{tabular}
        }

        \vspace{2pt} % 表格间的垂直间距

        \centering 
        % --- 第三个表格 ---
        {
        \begin{tabular}{@{}c | cccccc | cccccc@{}} 
        \toprule
        
        \multirow{2}{*}{\textbf{Methods}} & \multicolumn{6}{c|}{\textbf{RED-Trading-095}} & \multicolumn{6}{c}{\textbf{RED-Trading-0916}} \\
        
        & TB & DM & MS & FT & ACC & P90 & TB & DM & MS & FT & ACC & P90  \\
        
        \midrule
        
        % \longchainDS{}      & 68.18 & 68.18 & 59.09 & 89.39 & 59.09 & 26    & 68.85 & 68.85 & 60.65 & 95.08 & 60.67 & 34    \\
        % \longchainQwen      & 51.52 & 68.18 & 57.57 & 92.42 & 43.94 & 30    & 49.18 & 63.93 & 59.02 & 95.08 & 47.54 & 25    \\
        % \midrule
        % \modelRaw{}         & 90.91 & 90.91 & 66.67 & 83.33 & 53.03 (\textbf{$\downarrow$16.13}) & 7.8 (\textbf{$\times$3.06})   & 86.89 & 77.05 & 72.13 & 86.89 & 52.46 (\textbf{$-$}) & 6.98 (\textbf{$\times$3.27}) \\
        % \modelMan{}         & 89.39 & 98.48 & 78.79 & 95.45 & 65.15 (\textbf{$\uparrow$16.13}) & 10.66 (\textbf{$\times$2.26}) & 88.52 & 81.97 & 73.77 & 80.33 & 54.1 (\textbf{$-$})  & 11.12 (\textbf{$\times$2.41}) \\
        % \model{}            & 90.91 & 98.48 & 78.79 & 95.45 & 66.67 (\textbf{$\uparrow$16.13}) & 7.8 (\textbf{$\times$3.06})  & 90.16 & 81.97 & 77.05 & 86.89 & 59.02 (\textbf{$\uparrow$12.5}) & 7.74 (\textbf{$\times$3.15}) \\
        
        \longchainDS{}   & 67.74 & 67.74 & 22.58 & 96.77 & \textbf{22.58} & \textbf{24.00} & 41.67 & 41.67 & 20.83 & 83.33 & \textbf{20.83} & \textbf{29.00} \\
        \longchainQwen{} & 29.03 & 51.62 & 25.81 & 87.10 & \textbf{16.13} & \textbf{24.00} & 12.50 & 37.50 & 20.84 & 87.50 & \textbf{8.33}  & \textbf{24.00} \\
        \midrule
        % \modelRaw{}      & 83.87 & 90.32 & 61.29 & 90.32 & 38.71~(\textbf{$\uparrow$16.13}) & 7.82~(\textbf{$\times$3.07})  & 79.17 & 87.50 & 20.83 & 91.67 & 20.83~($-$) & 8.85~(\textbf{$\times$3.28})  \\
        \modelRaw{}     & 83.87 & 90.32 & 61.29 & 90.32 & \textbf{38.71} & \textbf{7.82}  & 79.17 & 87.50 & 20.83 & 91.67 & \textbf{20.83} & \textbf{8.85}  \\
        
        % \modelMan{}      & 83.87 & 77.42 & 54.84 & 93.55 & 38.71~(\textbf{$\uparrow$16.13}) & 10.60~(\textbf{$\times$2.26}) & 79.17 & 87.50 & 20.83 & 91.67 & 20.83~($-$) & 12.01~(\textbf{$\times$2.41}) \\
        \modelMan{}      & 83.87 & 77.42 & 54.84 & 93.55 & \textbf{38.71} &\textbf{10.60} & 79.17 & 87.50 & 20.83 & 91.67 & \textbf{20.83} & \textbf{12.01} \\
        
        % \model{}         & 83.87 & 77.42 & 58.06 & 83.87 & 38.71~(\textbf{$\uparrow$16.13}) & 7.82~(\textbf{$\times$3.07})  & 75.00 & 87.50 & 33.33 & 91.67 & 33.33~(\textbf{$\uparrow$12.50}) & 9.19~(\textbf{$\times$3.16})  \\
        \model{}         & 83.87 & 77.42 & 58.06 & 83.87 & \textbf{38.71} & \textbf{7.82}  & 75.00 & 87.50 & 33.33 & 91.67 & \textbf{33.33} & \textbf{9.19}  \\

        \rowcolor{blue!10}
        $\Delta$    &+16.13  &+9.68  &-12.9   &+35.48 &\textbf{+16.13} &\textbf{$\times$3.07}   &+33.33 &+45.83 &+12.5  &+8.37    &\textbf{+12.5}  &\textbf{$\times$3.16} \\
                
        \bottomrule
        
        \end{tabular}
        }
    % } 
    % \end{adjustbox}

\end{table*}

\subsubsection{Metrics}
\label{section:metrics}
We adopt the Execution Accuracy~(ACC) as the main metric and provide the accuracy of table selection~(TB), and component-level matching accuracies for each element of the DSL, namely DM~(dimension accuracy), MS~(measure accuracy) and FT~(filter accuracy). In addition, we include $P_{90}$, the 90th percentile of all request latencies, which is commonly used in enterprise scenarios, to evaluate the time-efficiency of \model{}. Consider a test set $\mathcal{D}_t = \{Q_1, Q_2, \cdots Q_m\}$ consisting of $m$ online user queries. Each query $Q_i \in \mathcal{D}_t$ is processed by the NL-to-DSL pipeline, and its inference latency $t_i$ is measured. Then these $m$ latency values are sorted in non-decreasing order to form a sequence $T = (t_{(1)}, t_{(2)}, \cdots t_{(m)})$, such that $t_{(j)} \leq t_{(j+1)}$ for all $j \in \{1, 2, \cdots m-1\}$ where $t_{(i)}$ represents the i-th smallest latency in the sequence of all. Thus, the metric $P_{90}$ can be formally defined as:
\begin{equation}
    P_{90} = t_{(\lceil m*0.9\rceil)}
\end{equation}

\subsubsection{Implementations}
\label{section:implemantations}
We evaluate two pipelines: the original long-chain baseline and our proposed shortcut \model{}. 
For long-chain methods, we derive two settings from the LLM workflow. The first, \longchainDS{}, involves three steps: (1) Query analysis using Qwen2.5-72b~\cite{yang2025qwen3}; (2) Data retrieval using deepseek-v3~\cite{liu2024deepseek} with BM25 and dense retrievers (tao-8k~\cite{tao8k}); and (3) DSL configuration using deepseek-v3 with BM25 and knowledge retrievers (tao-8k). The second setting, \longchainQwen{}, follows the same workflow but employs Qwen2.5-72b for both the data retrieval and DSL configuration steps. 
% 
% For our method, three settings \modelMan{}, \modelRaw{} and \model{} are derived, categorized by the encoding methods on user queries, with (1) manually labeled skeletons encoded by Qwen3-embedding-0.6B~\cite{zhang2025qwen3}, raw user queries encoded by Qwen3-embedding-0.6B, and raw user queries encoded by our entity-agnostic embedding model~(Qwen3-embedding-0.6B as the backbone, trained with the sentence-transformers library), respectively; (2) Qwen2.5-72b. Template-pool hyperparameters: For *-covered datasets, sim $\tau_s$=0.95, connectivity threshold=4, in-group top-k=2; For *-budget datasets, sim $\tau_s$=0.9, connectivity threshold=4, in-group top-k=3.
For our method, three settings \modelMan{}, \modelRaw{} and \model{} are derived, categorized by the encoding methods on user queries, with (1) manually labeled skeletons encoded by Qwen3-embedding-0.6B~\cite{zhang2025qwen3}, raw user queries encoded by Qwen3-embedding-0.6B, and raw user queries encoded by our entity-agnostic embedding model~(Qwen3-embedding-0.6B as the backbone, trained with the sentence-transformers library), respectively; (2) Qwen2.5-72b. Template-pool hyperparameters: For *-covered datasets, sim $\tau_s$=0.95, connectivity threshold=4, in-group top-k=2; For *-budget datasets, sim $\tau_s$=0.9, connectivity threshold=4, in-group top-k=3.

\subsubsection{Offline Cost}
In our evaluation using the industrial datasets \commerce{}, \community{}, and \trading{}, the skeleton construction phase requires 44.2s, 14.0s, and 10.3s, respectively, while the joint contrastive learning stage for these domains completes in approximately 3 minutes and 28 seconds on a single NVIDIA H100 GPU.
% \change{In our evaluation using the\rone{R1.C2} industrial datasets \commerce{}, \community{}, and \trading{}, the skeleton construction phase requires 44.2s, 14.0s, and 10.3s, respectively, while the joint contrastive learning stage for these domains completes in approximately 3 minutes and 28 seconds on a single NVIDIA H100 GPU.}

\subsection{Main Results~(RQ1)}
\subsubsection{The Long-chain Workflow Suffers from High Inference Latency} Consistent with our preliminary findings (Section~\ref{section:preliminary}), long-chain workflows, namely \longchainDS{} and \longchainQwen{}, exhibit consistently high latencies, exceeding 20s as illustrated in Table~\ref{tab:main-a-final}. On the \communityCovered{} dataset, this issue is more pronounced: the average P90 of long-chain methods reaches 29.5s, severely degrading the online user experience. We attribute this primarily to the multi-invocation nature of the LLM workflow, which is substantiated by the performance gap between \longchainDS{} and \longchainQwen{}; the latter achieves lower latency by substituting the reasoning-focused deepseek-v3 with the more inference-efficient Qwen2.5-72b for the DSL generation.

\subsubsection{\model{} Delivers Substantial Latency Reduction without Compromising Accuracy} As illustrated in Fig.~\ref{fig2:main}, \model{} delivers substantial latency reductions, significantly outperforming the leading long-chain competitor, \longchainDS{}, in terms of P90. On the \texttt{-0916} and \texttt{-095} datasets, the average P90 is reduced by 21.3s and 16.4s, respectively, highlighting the remarkable efficiency of our proposed shortcut method. These gains stem from two key advantages: a single-pass LLM invocation enabled by the shortcut architecture and a more concise prompt context, empowered by our knowledge incorporation strategy. While achieving this improvement in latency, \model{} still outperforms \longchainDS{} on execution accuracies by an average of 8.26, demonstrating superior performance-efficiency. 
% \change{A detailed error analysis\rtwo{R2.C4} is provided in Appendix~\ref{error}.}

% 15.48
% 11.65
% 5.14
\subsubsection{\model{} Excels in Configuring DSLs Using Historical References} As shown by the delta values quoted in Table~\ref{tab:main-a-final}, although individual metrics exhibit some volatility, the overall performance trend is positive, with the average accuracies of DM, MS, and FT increasing by 15.48, 11.65, and 5.14 points, respectively. This improvement is achieved by leveraging abundant mappings from the cache, all within a single generation step, in contrast to the long-chain solution. This demonstrates that the shortcut design enables \model{} to holistically configure the DSL structure with high precision in a single pass, bypassing the iterative generation process.

\subsubsection{\model{} Achieves Satisfactory Accuracy on Table Selection} According to our preliminary study in Section~\ref{section:preliminary}, the effort of selecting tables for long-chain solutions incurs significant overhead, which is an implicit challenge within the ``Data Retrieval'' step. As shown in Table~\ref{tab:main-a-final}, \model{} achieves an average accuracy of 85.99\% on table selection, surpassing the baseline by an average of 21.59 points. By utilizing table settings from historical DSLs with a voting strategy, \model{} can identify the correct table in most cases without relying on tedious LLM-based table selection.

\subsubsection{\model{} Demonstrates  Robustness against Cache Misses} To evaluate \model{} against cache misses, we investigate the performance of \model{} when \longchainDS{} is employed as a fallback mechanism. Experimental results demonstrate that even with the fallback mechanism engaged, \model{} maintains superior performance compared to the \longchainDS{} baseline, maintaining an average P90 of 25.46s, and reaching accuracies of 48.55, 60.47, and 23.08 on the evaluated datasets, respectively.

\begin{table}[h]
\caption{Performance comparison of RedParrot and ICL on \newspider{} and \newbird{}.}
\label{tab:performance-comparison}
\centering
\footnotesize
\renewcommand{\arraystretch}{0.8}
% First table for Spider-DSL
\begin{tabular}{ccccc}
\toprule
\multirow{2}{*}{\textbf{Method}} & \multicolumn{4}{c}{\textbf{Spider-DSL Acc(\%)}} \\ 
\cmidrule(l){2-5} 
& \textbf{Simple} & \textbf{Moderate} & \textbf{Challenging} & \textbf{Overall} \\
\midrule
ICL       & 53.9  & 42.2  & 47.4  & \textbf{47.9}          \\
\model{}   & 87.3  & 73.9  & 71.3  & \textbf{77.8}  \\  
\rowcolor{blue!10}
$\Delta$    & +33.4    &+31.7 &+23.9 &\textbf{+29.9}    \\
\bottomrule
\end{tabular}

\vspace{1pt} 

% Second table for BIRD-DSL
\begin{tabular}{ccccc}
\toprule
\multirow{2}{*}{\textbf{Method}} & \multicolumn{4}{c}{\textbf{BIRD-DSL Acc(\%)}} \\ 
\cmidrule(l){2-5} 
& \textbf{Simple} & \textbf{Moderate} & \textbf{Challenging} & \textbf{Overall} \\
\midrule 
ICL       & 26.9  & 25.4  & 25.2  & \textbf{25.8}          \\
\model{}   & 73.4  & 64.0  & 59.3  & \textbf{65.5}  \\ 

\rowcolor{blue!10}
$\Delta$    & +46.5    &+38.6 &+34.1 &\textbf{+39.7}    \\
\bottomrule
\end{tabular}

\end{table}

\subsection{Spider-DSL \& BIRD-DSL~(RQ2)}
\label{section:new-datasets}

To empirically evaluate the effectiveness of \model{} on public benchmarks, we adapt and extend two widely-used Text-to-SQL datasets: Spider and BIRD. Specifically, we leverage ICL to translate the query-SQL pair into our gold DSL, which serves as the new ground truth. It should be noted that the column in this DSL refers to the column in the physical dataset, whereas columns from the enterprise dataset require complex business post-processing.

The original Spider and BIRD lack sufficient semantically similar queries to effectively simulate historical references. To address this, we augment the data with three categories of generated queries: Simple, Moderate, and Challenging. Simple instances introduce linguistic variations by paraphrasing NL queries while keeping DSLs identical. Moderate and Challenging categories modify the DSL filter: Moderate by changing only the filter value, and Challenging by replacing the column-value pair. Generating these two categories requires semantic understandings of the original query and schema to introduce a valid new filter, followed by updating the NL query accordingly. After filtering out low-quality data based on the LLM-as-a-judge\cite{zheng2023judging} approach, we define the initially constructed gold DSLs as the template cache and use the remaining set of generated similar data as our test set. Finally, we constructed 769 test cases for \newspider{} and 1001 for \newbird{} from the gold DSLs.

We present the main experimental results in Table~\ref{tab:performance-comparison}, which compares the performance of \model{} against a standard DSL generation baseline without template caching (denoted as ICL). On \newspider{}, \model{} represents a substantial improvement of 29.9\% over the ICL baseline (47.9). The results on \newbird{}, a more challenging benchmark, are even more compelling. \model{} attains an accuracy of 65.5, outperforming the baseline (25.8) by a remarkable 39.7\%. The superiority of \model{} becomes more evident on the challenging benchmark, demonstrating its robustness in complex real-world scenarios.

\begin{table}[htbp]
    \caption{Ablation results of \model{}.}
    \label{tab:ablation} 
    \centering
    \renewcommand{\arraystretch}{0.9}
    \setlength{\tabcolsep}{3pt} % 极大地减少水平内边距

        \centering 
        % --- 第一个表格 ---
        {
        \begin{tabular}{@{}c | cccccc @{}} 
        \toprule
        
        \multirow{2}{*}{\textbf{Methods}}   & \multicolumn{6}{c}{\textbf{RED-Commerce-0916}} \\
        
        & TB     & DM     & MS   & FT    & ACC~(\%)  & P90~(s) \\
        \midrule
        \model{}   & 90.62 & 81.50 & 70.24 & 90.62 & 57.64 & 7.15 \\
        \midrule
        -encoder   & 88.47 & 79.09 & 67.56 & 92.22 & 51.74~(\textbf{$\downarrow$5.90}) & / \\
        % $\Delta$   & $-2.15$  & $-2.41$  & $-2.68$  & $+0.80$  & $-5.90$  & /  \\

        -knowledge  & 91.26 & 90.98 & 82.51 & 65.30 & 53.36~(\textbf{$\downarrow$4.28}) & / \\  
        
        -cache  & 70.25 & 80.97 & 51.48 & 95.98 & 45.84~(\textbf{$\downarrow$11.80}) & 25.00~(\textbf{$\times$3.50}) \\
        % $\Delta$   & $-20.37$  & $-0.53$  & $-18.76$ & $+4.56$ & $-11.80$ & $-20s$ \\
        
        \bottomrule
        \end{tabular}
        }

        \vspace{1pt}

        {
        \begin{tabular}{@{}c | cccccc @{}} 
        \toprule
        
        \multirow{2}{*}{\textbf{Methods}}   & \multicolumn{6}{c}{\textbf{RED-Community-0916}} \\
        
        & TB    & DM   & MS   & FT   & ACC~(\%)  & P90~(s) \\
        \midrule
        \model{}   & 90.16 & 81.97 & 77.05 & 86.89 & 59.02 & 7.74 \\
        \midrule
        -encoder   & 88.52 & 80.33 & 73.77 & 85.25 & 50.82~(\textbf{$\downarrow$8.20}) & / \\
        % $\Delta$   & $-1.64$  & $-1.64$  & $-3.28$ & $-1.64$  & $-8.20$  & / \\

        -knowledge  & 86.89 & 83.61 & 83.61 & 68.85 & 52.46~(\textbf{$\downarrow$6.56}) & / \\  
        
        -cache  & 68.85 & 68.85 & 60.65 & 95.08 & 60.67~(\textbf{$\uparrow$1.64}) & 34.00~(\textbf{$\times$4.39}) \\
        % $\Delta$   & $-21.31$  & $-13.12$  & $-16.40$ & $+8.19$ & $+1.64$ & $-29s$ \\

        \bottomrule
        \end{tabular}
        }

        \vspace{1pt}

        {
        \begin{tabular}{@{}c | cccccc @{}} 
        \toprule
        
        \multirow{2}{*}{\textbf{Methods}} & \multicolumn{6}{c}{\textbf{RED-Trading-0916}}\\
        
        & TB    & DM   & MS   & FT   & ACC~(\%)  & P90~(s) \\
        \midrule
        \model{}   & 75.00 & 87.50 & 33.33 & 91.67 & 33.33 & 7.74 \\
        \midrule
        -encoder   & 75.00 & 87.50 & 33.33 & 91.67 & 33.33~(\textbf{-}) & /  \\
        % $\Delta$   & $-0.00$  & $-0.00$  & $-0.00$ & $-0.00$  & $-0.00$  & /  \\

        -knowledge  & 91.67 & 75.00 & 95.83 & 68.85 & 37.50~(\textbf{$\uparrow$4.17}) & /  \\ 
        
        -cache  & 41.67 & 41.67 & 20.83 & 83.33 & 20.83~(\textbf{$\downarrow$12.50}) & 29.00~(\textbf{$\times$3.75})     \\
        % $\Delta$   & $-33.34$  & $-45.84$  & $-12.50$ & $-8.34$ & $-12.5$ & $-24s$ \\
        
        \bottomrule
        \end{tabular}
        }

\end{table}

\subsection{Ablation Study~(RQ3)}
We conduct an ablation study regarding our entity-agnostic encoder and knowledge incorporation with results presented in Table~\ref{tab:ablation}.

\textbf{Entity-agnostic encoder.}
We replaced our entity-agnostic encoder with a two-step baseline approach that first extracts the skeleton via NER and then encodes it for vector retrieval using a SOTA embedding model of equivalent scale. This substitution caused a significant drop in execution accuracy (denoted ACC) on both \commerceCovered{} and \communityCovered{}, while \tradingCovered{} remained unchanged.
This can be attributed to our entity-agnostic encoder's ability to effectively learn skeleton patterns, which allows it to distinguish distractor words more precisely than the NER-based method.
% \rone{R1.W3}\change{We replaced our entity-agnostic encoder with a two-step baseline approach that first extracts the skeleton via NER and then encodes it for vector retrieval using a SOTA embedding model of equivalent scale.} This substitution caused a significant drop in execution accuracy (denoted ACC) on both \commerceCovered{} and \communityCovered{}, while \tradingCovered{} remained unchanged.
% This can be attributed to our entity-agnostic encoder's ability to effectively learn skeleton patterns, which allows it to distinguish distractor words more precisely than the NER-based method.

\textbf{Knowledge incorporation.}
Removing the knowledge-retrieval module from the short pipeline resulted in a substantial decrease in ACC on both \commerceCovered{} and \communityCovered{}, while the accuracy on \tradingCovered{} unexpectedly shown a marginal increase. This counter-intuitive result on the \tradingCovered{} dataset is likely attributed to its limited scale, as detailed in Table~\ref{tab:datasets}. However, the significant drop on the larger datasets confirms the retrieval component's value: it supplies crucial domain knowledge, enabling the short pipeline to handle new information in queries absent from historical (skeleton, DSL) pairs.

\textbf{Skeleton Cache.}
We replaced the short pipeline accelerated by the skeleton cache with the long pipeline, which led to ACC decreases in \commerceCovered{} and \tradingCovered{}, while the accuracy on \communityCovered{} dropped slightly. For each of the three datasets, the runtime increased by at least a factor of three. This demonstrates that our short pipeline maintains or even improves accuracy while substantially reducing latency.

\subsection{Effectiveness of the Entity-agnostic Model~(RQ4)}
\label{section:entity-agnostic-effectiveness}
\begin{figure}[htbp]
\centerline{\includegraphics[width=1\columnwidth]{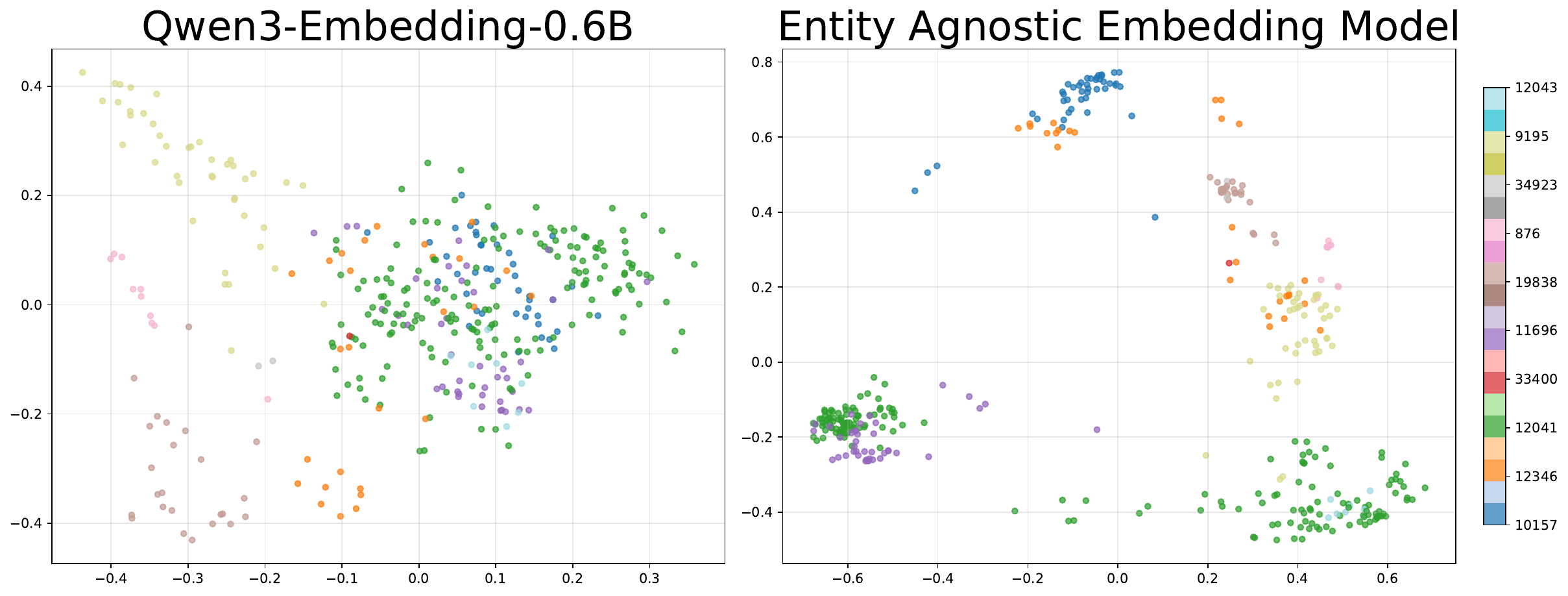}}
\caption{Embedding Space Comparison via PCA.}
\label{fig:entity-agnostic-effectiveness}
\end{figure}

In this section, we evaluate the quality of the learned representations by visualizing them using PCA. As illustrated in Fig.~\ref{fig:entity-agnostic-effectiveness}, the baseline Qwen3-Embedding-0.6B model (left) exhibits significant overlap between different entity classes (indicated by color), resulting in a poorly defined cluster structure. Conversely, the Entity Agnostic Model (right) produces highly discriminative representations, forming compact and well-separated clusters for each entity. This result illustrates the superior class separability achieved by our entity-agnostic approach, thereby validating its effectiveness in learning discriminative and generalizable entity representations.

% \change{In this section, we evaluate the quality of the learned representations by visualizing them using PCA. As illustrated in Fig.~\ref{fig:entity-agnostic-effectiveness}, the baseline Qwen3-Embedding-0.6B model (left) exhibits significant overlap between different entity classes (indicated by color), resulting in a poorly defined cluster structure. Conversely, the Entity Agnostic Model (right) produces highly discriminative representations, forming compact and well-separated clusters for each entity. This result illustrates the superior class separability achieved by our entity-agnostic approach, thereby validating its effectiveness in learning discriminative and generalizable entity representations.}

\begin{table}[h]
\caption{Hit rate performance comparison.}
\label{recall}
\centering
\footnotesize
\renewcommand{\arraystretch}{0.775}
\begin{tabular}{c|c|c|c} % 第3列（现在是ACC）保持加粗
\toprule
\textbf{Dataset} & \textbf{Embedding Model} & \textbf{HR@5} & \textbf{FHR@5} \\ 
\midrule
\multirow{2}{*}{\textbf{Commerce}}  & Qwen3-embedding-0.6B             & 85.79  & 56.84 \\
                           & Entity-agnostic model & 93.57   & 75.34 \\
                           & \cellcolor{blue!10} $\Delta$    & \cellcolor{blue!10} \textbf{7.78}  & \cellcolor{blue!10} \textbf{19.4}  \\ 
                           % ({\color{green!60!black}$\uparrow$7.78})
\midrule
\multirow{2}{*}{\textbf{Community}} & Qwen3-embedding-0.6B             & 95.08  & 81.97 \\
                           & Entity-agnostic model & 100  & 100    \\
                           & \cellcolor{blue!10} $\Delta$    & \cellcolor{blue!10} \textbf{4.92}  & \cellcolor{blue!10} \textbf{18.03} \\ 
\midrule
\multirow{2}{*}{\textbf{Trading}}   & Qwen3-embedding-0.6B             & 100  & 100    \\
                           & Entity-agnostic model & 100  & 100    \\
                           & \cellcolor{blue!10} $\Delta$    & \cellcolor{blue!10} -    & \cellcolor{blue!10} -      \\ 
\bottomrule
\end{tabular}

\end{table}

\begin{figure}[ht] 
\centering 
\includegraphics[width=1\columnwidth]{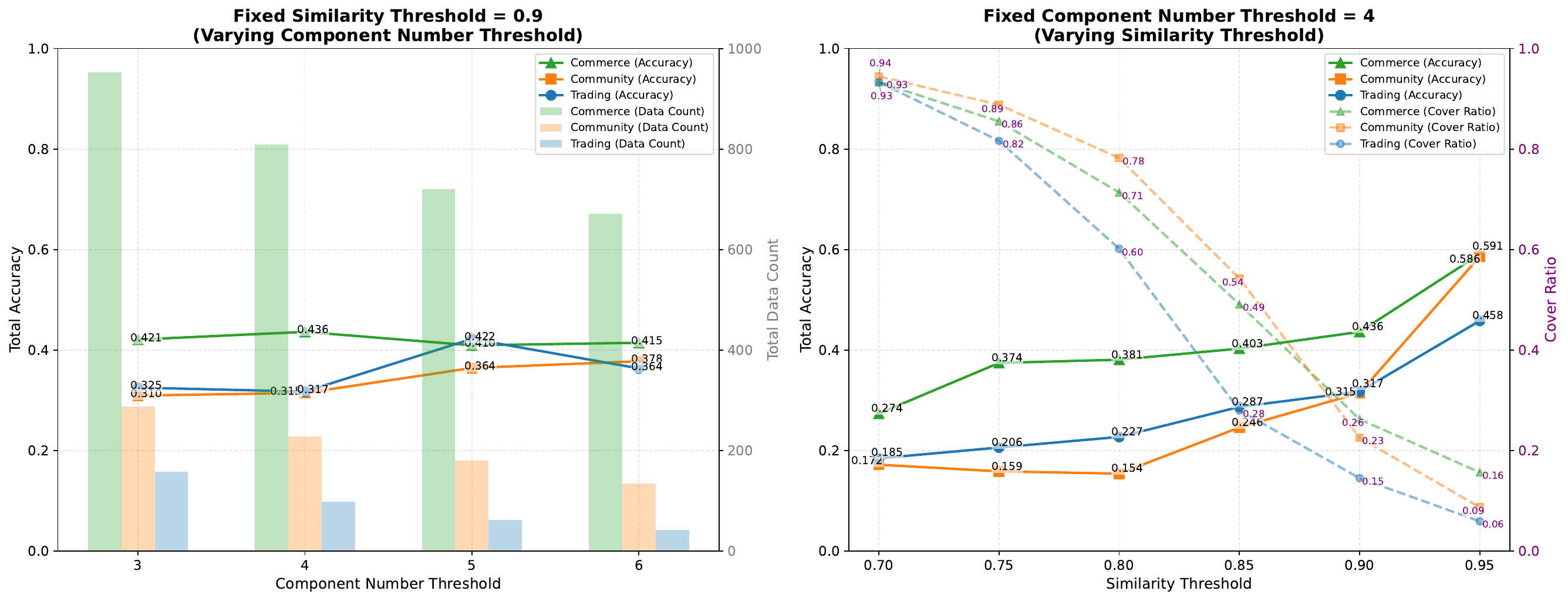}
\caption{Sensitivity Analysis of Similarity Threshold and Component Number Threshold.}
\label{fig:sensitivity}
\end{figure}

Additional experiments were conducted to verify the effectiveness of our embedding model. 1) To evaluate the system's ability to handle multi-intent queries, we employ Hit Rate@5 (HR@5) to measure the probability of retrieving at least one relevant target, and Full Hit Rate@5 (FHR@5) to assess the more stringent requirement of capturing all ground-truth targets within the top-5 results. Across three latest enterprise datasets, our entity-agnostic model outperforms the equivalent-scale SOTA Qwen3-embedding-0.6B by an average of 4.23 and 12.47 points in HR@5 and FHR@5, respectively. 2) We perform a sensitivity analysis of the hyperparameters used by the system. As shown in Fig.~\ref{fig:sensitivity}, we set the similarity threshold $\tau_s=0.9$ to ensure semantic precision in cache matching without compromising problem coverage ratio. The connectivity cutoff is implemented as a vertex degree threshold ($\tau_c>4$) within our similarity-based graph, which identifies representative skeletons and filters out idiosyncratic noise. Overall, the system exhibits low sensitivity to these hyperparameters, as their fluctuations align with expected performance trade-offs.

% \change{Additional experiments were conducted to verify the effectiveness of our embedding model. 1)\rone{R1.W1, R1.C5} To evaluate the system's ability to handle multi-intent queries, we employ Hit Rate@5 (HR@5) to measure the probability of retrieving at least one relevant target, and Full Hit Rate@5 (FHR@5) to assess the more stringent requirement of capturing all ground-truth targets within the top-5 results. Across three latest enterprise datasets, our entity-agnostic model outperforms the equivalent-scale SOTA Qwen3-embedding-0.6B by an average of 4.23 and 12.47 points in HR@5 and FHR@5, respectively. 2) We \rone{R1.C1}perform a sensitivity analysis of the hyperparameters used by the system. As shown in Fig.~\ref{fig:sensitivity}, we set the similarity threshold $\tau_s=0.9$ to ensure semantic precision in cache matching without compromising problem coverage ratio. The connectivity cutoff is implemented as a vertex degree threshold ($\tau_c>4$) within our similarity-based graph, which identifies representative skeletons and filters out idiosyncratic noise. Overall, the system exhibits low sensitivity to these hyperparameters, as their fluctuations align with expected performance trade-offs.}

\subsection{Analysis of the Cache Updating Strategies~(RQ5)}
\label{sec:cache_update}

We provide two strategies for updating the skeleton cache: a full rebuild and an incremental update. The former baseline triggers a comprehensive global rebuild whenever the volume of newly accumulated queries reaches a $10\%$ threshold (approximately 580 instances relative to the initial pool of 5,854). While this strategy updates the cache by re-processing the entire dataset, the resulting overhead becomes prohibitive at larger scales. To mitigate this, we introduce an online incremental update strategy that identifies representative candidates through connectivity-based filtering of incoming query clusters. 
A candidate is cached only if its similarity $s$ to existing templates meets specific criteria. Specifically, a similarity of $s > 0.95$ identifies high-confidence patterns for reinforcement, while $s < 0.9$ indicates novel structural diversity that warrants inclusion. 
These high thresholds are chosen because the similarity distribution between new and historical queries at Xiaohongshu is highly concentrated.

% \change{\rthree{R1.C6, R2.C2, R3.W1}We provide two strategies for updating the skeleton cache: a full rebuild and an incremental update. The former baseline triggers a comprehensive global rebuild whenever the volume of newly accumulated queries reaches a $10\%$ threshold (approximately 580 instances relative to the initial pool of 5,854). While this strategy updates the cache by re-processing the entire dataset, the resulting overhead becomes prohibitive at larger scales. To mitigate this, we introduce an online incremental update strategy that identifies representative candidates through connectivity-based filtering of incoming query clusters. 
% % 
% A candidate is cached only if its similarity $s$ to existing templates meets specific criteria. Specifically, a similarity of $s > 0.95$ identifies high-confidence patterns for reinforcement, while $s < 0.9$ indicates novel structural diversity that warrants inclusion. 
% % 
% These high thresholds are chosen because the similarity distribution between new and historical queries at Xiaohongshu is highly concentrated.}

\begin{table}[h]
\caption{Comparison among cache updating strategies.}
\label{recall}
\centering
\footnotesize
\renewcommand{\arraystretch}{0.85}
\begin{tabular}{@{}c|c|c|c|c@{}} % 第3列（现在是ACC）保持加粗
\toprule
\textbf{Dataset} & \textbf{Strategy} & \textbf{ACC (\%)} & \textbf{Cov. (\%)}   & \textbf{Overhead (s)}  \\ 
\midrule
\multirow{3}{*}{\textbf{Commerce}}  & \textbf{N/A}                & 46.55     & 22.13        & 50.11    \\
                                    & \textbf{Full Rebuild}         & 41.34     & 22.13      & 48.7    \\
                                    & \textbf{Incremental}           & 51.72       & 22.13   & 9.48 \textbf{($\times 5.13$)}    \\ 
\midrule
\multirow{3}{*}{\textbf{Community}}  & \textbf{N/A}               & 64.28     & 12.17   & 15.66    \\
                                    & \textbf{Full Rebuild}         & 58.26     & 14.71  & 16.61    \\
                                    & \textbf{Incremental}           & 60.0       & 13.04    & 4.93  \textbf{($\times3.36$)}  \\ 
\midrule
\multirow{3}{*}{\textbf{Trading}}   & \textbf{N/A}                & 76.19     & 17.79    & 14.38    \\
                                    & \textbf{Full Rebuild}         & 80.02     & 17.79  & 19.62        \\
                                    & \textbf{Incremental}           & 76.19       & 17.79   & 7.47 \textbf{($\times2.62$)}       \\

\bottomrule
\end{tabular}
\end{table}

Experimental results (Table~\ref{recall}) reveal that while accuracy fluctuates across datasets due to the high similarity between incremental and historical queries, the incremental update strategy delivers a transformative reduction in computational latency. Achieving an average $3.7\times$ speedup, this approach effectively sidesteps the prohibitive overhead of a global rebuild, particularly in the scale-heavy \commerce{} domain where it realizes a $5.13\times$ efficiency gain. 
Although the full rebuild occasionally underperforms—likely because exhaustive re-processing introduces global noise—the incremental strategy maintains competitive precision and superior structural coverage. 
The incremental strategy effectively preserves the precision of the DSL pool while ensuring the feasibility of continuous, low-overhead cache evolution.

% \change{Experimental results (Table~\ref{recall}) reveal that while accuracy fluctuates across datasets due to the high similarity between incremental and historical queries, the incremental update strategy delivers a transformative reduction in computational latency. Achieving an average $3.7\times$ speedup, this approach effectively sidesteps the prohibitive overhead of a global rebuild, particularly in the scale-heavy \commerce{} domain where it realizes a $5.13\times$ efficiency gain. 
% % 
% Although the full rebuild occasionally underperforms—likely because exhaustive re-processing introduces global noise—the incremental strategy maintains competitive precision and superior structural coverage. 
% % 
% The incremental strategy effectively preserves the precision of the DSL pool while ensuring the feasibility of continuous, low-overhead cache evolution.}

\subsection{Error Analysis~(RQ1)}

\begin{figure}[htbp] 
\centering 
\includegraphics[width=\columnwidth]{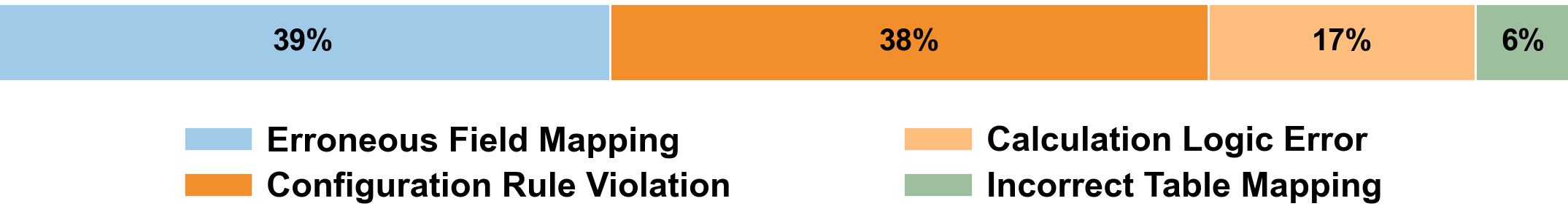}
\caption{Distribution of execution error categories in NL-to-DSL generation.}
\label{fig:error_distribution}
\end{figure}

In this section, we incorporate an error analysis based on failure cases evaluated. Our analysis categorizes the execution errors into four types: Erroneous Field Mapping (39\%) and Incorrect Table Mapping (6\%), which often stem from the retrieval noise mentioned; Configuration Rule Violation (38\%), arising from atypical business fields; and Calculation Logic Errors (17\%), which primarily affect complex, multi-intent queries. For long or structurally intricate queries that challenge single-pass logic, the system relies on the aforementioned fallback mechanism.

% \change{In this section, we incorporate an error analysis\rtwo{R2.C4} based on failure cases evaluated. Our analysis categorizes the execution errors into four types: Erroneous Field Mapping\rtwo{R2.C3} (39\%) and Incorrect Table Mapping (6\%), which often stem from the retrieval noise mentioned; Configuration Rule Violation (38\%), arising from atypical business fields; and Calculation Logic Errors (17\%), which primarily affect complex, multi-intent queries. For long or structurally intricate queries that challenge single-pass logic, the system relies on the aforementioned fallback mechanism.}

\subsection{Depolyment Architecture}

The deployment architecture of \model{} is structured around a dual-path execution strategy (Short-chain and Long-chain) orchestrated by an internal workflow engine. To manage the lifecycle of query skeletons, the system employs an offline-to-online pipeline: high-quality query templates are pre-constructed through K-means clustering and stored in a Milvus-backed vectorized repository, while the online serving layer utilizes a Query Erase module to extract structural patterns from raw NL inputs. All processing nodes, from intent parsing to DSL generation, are encapsulated within a modular microservice framework, enabling asynchronous performance monitoring and automatic fallback from short-chain template matching to the heterogeneous RAG-driven long-chain when similarity thresholds are not met.

% \change{The deployment architecture of \model{}\rtwo{R2.W1} is structured around a dual-path execution strategy (Short-chain and Long-chain) orchestrated by an internal workflow engine. To manage the lifecycle of query skeletons, the system employs an offline-to-online pipeline: high-quality query templates are pre-constructed through K-means clustering and stored in a Milvus-backed vectorized repository, while the online serving layer utilizes a Query Erase module to extract structural patterns from raw NL inputs. All processing nodes, from intent parsing to DSL generation, are encapsulated within a modular microservice framework, enabling asynchronous performance monitoring and automatic fallback from short-chain template matching to the heterogeneous RAG-driven long-chain when similarity thresholds are not met.}
\section{Related Works}
% data collection, storage, preparation, analysis, and visualization
% Business analytics is a technology-driven process that facilitates data-informed decision-making, primarily through data analysis and data visualization~\cite{quamar2020conversational}. The recent proliferation of LLMs has significantly impacted the data analysis stage by lowering the technical barrier, empowering non-expert users to interact with complex databases through NL\cite{zhu2024chat2query, weng2025datalab}. 

The recent proliferation of LLMs has significantly lowered the technical barrier for data analysis, empowering non-expert users to conduct business analytics and interact with complex databases through natural language~\cite{zhu2024chat2query,weng2025datalab}.

Within the domain of enterprise business analytics, where strategic decisions are critically contingent on data analysis, a semantic parsing paradigm is widely adopted\cite{su2024tablegpt2}. This methodology involves first translating a NL query into a structured intermediate representation defined by a DSL\cite{wang2023grammar}. By explicitly modeling the query’s analytical intent, this DSL specification acts as a canonical source of truth, enabling the robust and deterministic generation of outputs for downstream tasks such as NL2SQL\cite{liu2024survey} and NL2VIS\cite{chen2024viseval}. DataLab\cite{weng2025datalab} presents a unified business analytics framework that leverages an intermediate DSL to bridge NL queries and executable code like SQL and Vega-Lite. TableGPT2\cite{su2024tablegpt2} introduces Grammar Prompting to improve the generation of highly structured DSLs from few-shot examples. This approach reframes the DSL from a static output format into a dynamic, learnable constraint, enabling the model to better generalize to complex structured languages.

In parallel to methods centered on intermediate DSLs, researchers have also addressed other practical challenges in deploying LLM-based business analytics systems. Chat2Query\cite{zhu2024chat2query} was developed as a zero-shot system that empowers users to perform exploratory data analysis by generating not only SQL queries but also suitable visualizations from NL prompts. Chat2Data\cite{zhao2024chat2data} employs a three-layer architecture incorporating RAG to infuse domain knowledge, vector databases to reduce costly LLM interactions, and a pipeline agent to decompose complex tasks into manageable subtasks, thereby enhancing reliability and accuracy. To overcome the token limitations posed by large database schemas, ChatBI\cite{lian2024chatbi} innovatively employs database view technology and a smaller machine learning model to first prune the schema to a relevant subset before providing it to the LLM for SQL generation.

% However, prior works often rely on multi-step workflows requiring sequential LLM invocations, which inherently leads to error propagation and cumulative inference latency.
However, prior work often relies on multi-step workflows involving sequential LLM invocations, leading to error propagation and accumulated inference latency.
To address these limitations, we propose \model{}, which is designed to ensure both high task performance and low inference time.

% DataLab description
% DataLab\cite{weng2025datalab} introduces a unified platform that integrates a multi-agent framework within a computational notebook which supports inter-agent communication and incorporates domain-specific knowledge. 

% Why DSL?
% This intermediate representation(DSL) serves a crucial function: it decouples the complex task of natural language understanding from the deterministic generation of target languages like SQL and Vega-Lite. 

\section{Conclusion}

We presented \model{}, a novel, cache-based shortcut framework for NL-to-DSL translation that meets the high-performance and robustness demands of enterprise applications. \model{} focuses on constructing a historical skeleton cache and employing an entity-agnostic contrastive learning encoder to retrieve (skeleton, DSL) pairs from the cache. We further enhanced generation quality by employing an RAG method to augment the context with multi-level heterogeneous knowledge. Experiments show that our shortcut significantly accelerates the pipeline without loss of accuracy.
To validate the generality of our approach, we made another key contribution by converting the widely used Text-to-SQL datasets Spider and BIRD into the NL-to-DSL setting and achieving exceptional results on both datasets.
The code, datasets, and detailed supplementary appendix are available at \url{https://github.com/TommyIsNotHere/RedParrot}

% Our proposed shortcut \model{} introduces a new paradigm for NL-to-DSL and the first NL-to-DSL enterprise datasets offers a foundational resource for advancing future research in this field.

\section*{Acknowledgement}

% We hereby confirm that the entire work presented in this paper, including all content, analysis, and conclusions, is the original work of the human authors. While the core research, analysis, and conclusions are entirely the work of the human authors, generative AI was consulted for the limited purpose of improving the grammar and readability of the text.

This work was supported by the Pioneer R\&D Program of Zhejiang (No.2024C01021), the \enquote{Leading Talent of Technological Innovation Program} (No.2023R5214) of Zhejiang Province, and the collaborative project between Xiaohongshu and Zhejiang University.

\clearpage
\bibliographystyle{IEEEtran}
\bibliography{main}

\begin{thebibliography}{10}
\providecommand{\url}[1]{#1}
\csname url@samestyle\endcsname
\providecommand{\newblock}{\relax}
\providecommand{\bibinfo}[2]{#2}
\providecommand{\BIBentrySTDinterwordspacing}{\spaceskip=0pt\relax}
\providecommand{\BIBentryALTinterwordstretchfactor}{4}
\providecommand{\BIBentryALTinterwordspacing}{\spaceskip=\fontdimen2\font plus
\BIBentryALTinterwordstretchfactor\fontdimen3\font minus \fontdimen4\font\relax}
\providecommand{\BIBforeignlanguage}[2]{{%
\expandafter\ifx\csname l@#1\endcsname\relax
\typeout{** WARNING: IEEEtran.bst: No hyphenation pattern has been}%
\typeout{** loaded for the language `#1'. Using the pattern for}%
\typeout{** the default language instead.}%
\else
\language=\csname l@#1\endcsname
\fi
#2}}
\providecommand{\BIBdecl}{\relax}
\BIBdecl

\bibitem{zhu2024chat2query}
J.-P. Zhu, P.~Cai, B.~Niu, Z.~Ni, K.~Xu, J.~Huang, J.~Wan, S.~Ma, B.~Wang, D.~Zhang \emph{et~al.}, ``Chat2query: A zero-shot automatic exploratory data analysis system with large language models,'' in \emph{2024 IEEE 40th International Conference on Data Engineering (ICDE)}.\hskip 1em plus 0.5em minus 0.4em\relax IEEE, 2024, pp. 5429--5432.

\bibitem{weng2025datalab}
L.~Weng, Y.~Tang, Y.~Feng, Z.~Chang, R.~Chen, H.~Feng, C.~Hou, D.~Huang, Y.~Li, H.~Rao \emph{et~al.}, ``Datalab: A unified platform for llm-powered business intelligence,'' in \emph{2025 IEEE 41st International Conference on Data Engineering (ICDE)}.\hskip 1em plus 0.5em minus 0.4em\relax IEEE, 2025, pp. 4346--4359.

\bibitem{cui2025tablecopilot}
L.~Cui, G.~Jiang, H.~Li, K.~Chen, L.~Shou, and G.~Chen, ``Tablecopilot: A table assistant empowered by natural language conditional table discovery,'' \emph{arXiv preprint arXiv:2507.08283}, 2025.

\bibitem{yuan2025cogsql}
H.~Yuan, X.~Tang, K.~Chen, L.~Shou, G.~Chen, and H.~Li, ``Cogsql: A cognitive framework for enhancing large language models in text-to-sql translation,'' in \emph{Proceedings of the AAAI Conference on Artificial Intelligence}, vol.~39, no.~24, 2025, pp. 25\,778--25\,786.

\bibitem{gao2024text}
D.~Gao, H.~Wang, Y.~Li, X.~Sun, Y.~Qian, B.~Ding, and J.~Zhou, ``Text-to-sql empowered by large language models: A benchmark evaluation,'' \emph{Proceedings of the VLDB Endowment}, vol.~17, no.~5, pp. 1132--1145, 2024.

\bibitem{peng2024large}
W.~Peng, G.~Li, Y.~Jiang, Z.~Wang, D.~Ou, X.~Zeng, D.~Xu, T.~Xu, and E.~Chen, ``Large language model based long-tail query rewriting in taobao search,'' in \emph{Companion Proceedings of the ACM Web Conference 2024}, 2024, pp. 20--28.

\bibitem{balaka2025pneuma}
M.~I.~L. Balaka, D.~Alexander, Q.~Wang, Y.~Gong, A.~Krisnadhi, and R.~Castro~Fernandez, ``Pneuma: Leveraging llms for tabular data representation and retrieval in an end-to-end system,'' \emph{Proceedings of the ACM on Management of Data}, vol.~3, no.~3, pp. 1--28, 2025.

\bibitem{xu2024kcmf}
Y.~Xu, H.~Li, K.~Chen, and L.~Shou, ``Kcmf: A knowledge-compliant framework for schema and entity matching with fine-tuning-free llms,'' \emph{arXiv preprint arXiv:2410.12480}, 2024.

\bibitem{wang2023solo}
Q.~Wang and R.~Castro~Fernandez, ``Solo: Data discovery using natural language questions via a self-supervised approach,'' \emph{Proceedings of the ACM on Management of Data}, vol.~1, no.~4, pp. 1--27, 2023.

\bibitem{hu2025lakevisage}
Y.~Hu, J.~Wang, and S.~Rahman, ``Lakevisage: Towards scalable, flexible and interactive visualization recommendation for data discovery over data lakes,'' \emph{arXiv preprint arXiv:2504.02150}, 2025.

\bibitem{lewis2020retrieval}
P.~Lewis, E.~Perez, A.~Piktus, F.~Petroni, V.~Karpukhin, N.~Goyal, H.~K{\"u}ttler, M.~Lewis, W.-t. Yih, T.~Rockt{\"a}schel \emph{et~al.}, ``Retrieval-augmented generation for knowledge-intensive nlp tasks,'' \emph{Advances in neural information processing systems}, vol.~33, pp. 9459--9474, 2020.

\bibitem{yu2018spider}
T.~Yu, R.~Zhang, K.~Yang, M.~Yasunaga, D.~Wang, Z.~Li, J.~Ma, I.~Li, Q.~Yao, S.~Roman \emph{et~al.}, ``Spider: A large-scale human-labeled dataset for complex and cross-domain semantic parsing and text-to-sql task,'' in \emph{Proceedings of the 2018 Conference on Empirical Methods in Natural Language Processing}, 2018, pp. 3911--3921.

\bibitem{li2023can}
J.~Li, B.~Hui, G.~Qu, J.~Yang, B.~Li, B.~Li, B.~Wang, B.~Qin, R.~Geng, N.~Huo \emph{et~al.}, ``Can llm already serve as a database interface? a big bench for large-scale database grounded text-to-sqls,'' \emph{Advances in Neural Information Processing Systems}, vol.~36, pp. 42\,330--42\,357, 2023.

\bibitem{chen2016realtime}
G.~J. Chen, J.~L. Wiener, S.~Iyer, A.~Jaiswal, R.~Lei, N.~Simha, W.~Wang, K.~Wilfong, T.~Williamson, and S.~Yilmaz, ``Realtime data processing at facebook,'' in \emph{Proceedings of the 2016 International Conference on Management of Data}, 2016, pp. 1087--1098.

\bibitem{gupta2016mesa}
A.~Gupta, F.~Yang, J.~Govig, A.~Kirsch, K.~Chan, K.~Lai, S.~Wu, S.~Dhoot, A.~R. Kumar, A.~Agiwal \emph{et~al.}, ``Mesa: A geo-replicated online data warehouse for google's advertising system,'' \emph{Communications of the ACM}, vol.~59, no.~7, pp. 117--125, 2016.

\bibitem{begoli2018apache}
E.~Begoli, J.~Camacho-Rodr{\'\i}guez, J.~Hyde, M.~J. Mior, and D.~Lemire, ``Apache calcite: A foundational framework for optimized query processing over heterogeneous data sources,'' in \emph{Proceedings of the 2018 International Conference on Management of Data}, 2018, pp. 221--230.

\bibitem{siddiqui2021sketching}
T.~Siddiqui, P.~Luh, Z.~Wang, K.~Karahalios, and A.~G. Parameswaran, ``From sketching to natural language: Expressive visual querying for accelerating insight,'' \emph{ACM SIGMOD Record}, vol.~50, no.~1, pp. 51--58, 2021.

\bibitem{robertson2009probabilistic}
S.~Robertson, H.~Zaragoza \emph{et~al.}, ``The probabilistic relevance framework: Bm25 and beyond,'' \emph{Foundations and Trends{\textregistered} in Information Retrieval}, vol.~3, no.~4, pp. 333--389, 2009.

\bibitem{platzer2005vector}
C.~Platzer and S.~Dustdar, ``A vector space search engine for web services,'' in \emph{Third European Conference on Web Services (ECOWS'05)}.\hskip 1em plus 0.5em minus 0.4em\relax IEEE, 2005, pp. 9--pp.

\bibitem{HanLP}
\BIBentryALTinterwordspacing
``Hanlp.'' [Online]. Available: \url{https://github.com/hankcs/HanLP}
\BIBentrySTDinterwordspacing

\bibitem{ahmed2020k}
M.~Ahmed, R.~Seraj, and S.~M.~S. Islam, ``The k-means algorithm: A comprehensive survey and performance evaluation,'' \emph{Electronics}, vol.~9, no.~8, p. 1295, 2020.

\bibitem{khosla2020supervised}
P.~Khosla, P.~Teterwak, C.~Wang, A.~Sarna, Y.~Tian, P.~Isola, A.~Maschinot, C.~Liu, and D.~Krishnan, ``Supervised contrastive learning,'' \emph{Advances in neural information processing systems}, vol.~33, pp. 18\,661--18\,673, 2020.

\bibitem{mai2024learning}
C.~Mai, R.-e. Tal, and T.~Mohamed, ``Learning metadata-agnostic representations for text-to-sql in-context example selection,'' \emph{arXiv preprint arXiv:2410.14049}, 2024.

\bibitem{yang2025qwen3}
A.~Yang, A.~Li, B.~Yang, B.~Zhang, B.~Hui, B.~Zheng, B.~Yu, C.~Gao, C.~Huang, C.~Lv \emph{et~al.}, ``Qwen3 technical report,'' \emph{arXiv preprint arXiv:2505.09388}, 2025.

\bibitem{liu2024deepseek}
A.~Liu, B.~Feng, B.~Xue, B.~Wang, B.~Wu, C.~Lu, C.~Zhao, C.~Deng, C.~Zhang, C.~Ruan \emph{et~al.}, ``Deepseek-v3 technical report,'' \emph{arXiv preprint arXiv:2412.19437}, 2024.

\bibitem{tao8k}
\BIBentryALTinterwordspacing
``tao-8k.'' [Online]. Available: \url{https://huggingface.co/Amu/tao-8k}
\BIBentrySTDinterwordspacing

\bibitem{zhang2025qwen3}
Y.~Zhang, M.~Li, D.~Long, X.~Zhang, H.~Lin, B.~Yang, P.~Xie, A.~Yang, D.~Liu, J.~Lin \emph{et~al.}, ``Qwen3 embedding: Advancing text embedding and reranking through foundation models,'' \emph{arXiv preprint arXiv:2506.05176}, 2025.

\bibitem{zheng2023judging}
L.~Zheng, W.-L. Chiang, Y.~Sheng, S.~Zhuang, Z.~Wu, Y.~Zhuang, Z.~Lin, Z.~Li, D.~Li, E.~Xing \emph{et~al.}, ``Judging llm-as-a-judge with mt-bench and chatbot arena,'' \emph{Advances in neural information processing systems}, vol.~36, pp. 46\,595--46\,623, 2023.

\bibitem{su2024tablegpt2}
A.~Su, A.~Wang, C.~Ye, C.~Zhou, G.~Zhang, G.~Chen, G.~Zhu, H.~Wang, H.~Xu, H.~Chen \emph{et~al.}, ``Tablegpt2: A large multimodal model with tabular data integration,'' \emph{arXiv preprint arXiv:2411.02059}, 2024.

\bibitem{wang2023grammar}
B.~Wang, Z.~Wang, X.~Wang, Y.~Cao, R.~A~Saurous, and Y.~Kim, ``Grammar prompting for domain-specific language generation with large language models,'' \emph{Advances in Neural Information Processing Systems}, vol.~36, pp. 65\,030--65\,055, 2023.

\bibitem{liu2024survey}
X.~Liu, S.~Shen, B.~Li, P.~Ma, R.~Jiang, Y.~Zhang, J.~Fan, G.~Li, N.~Tang, and Y.~Luo, ``A survey of nl2sql with large language models: Where are we, and where are we going?'' \emph{arXiv preprint arXiv:2408.05109}, 2024.

\bibitem{chen2024viseval}
N.~Chen, Y.~Zhang, J.~Xu, K.~Ren, and Y.~Yang, ``Viseval: A benchmark for data visualization in the era of large language models,'' \emph{IEEE Transactions on Visualization and Computer Graphics}, 2024.

\bibitem{zhao2024chat2data}
X.~Zhao, X.~Zhou, and G.~Li, ``Chat2data: An interactive data analysis system with rag, vector databases and llms,'' \emph{Proceedings of the VLDB Endowment}, vol.~17, no.~12, pp. 4481--4484, 2024.

\bibitem{lian2024chatbi}
J.~Lian, X.~Liu, Y.~Shao, Y.~Dong, M.~Wang, Z.~Wei, T.~Wan, M.~Dong, and H.~Yan, ``Chatbi: Towards natural language to complex business intelligence sql,'' \emph{arXiv preprint arXiv:2405.00527}, 2024.

\end{thebibliography}

% \clearpage      
% \input{content/appendix}

\end{document}